\documentclass[final,3p,times,onecolumn,number,review,reviewsort&compress]{elsarticle}
\usepackage{amssymb}
\usepackage{upgreek}
\usepackage{hyperref}
\usepackage{makecell}
\usepackage{multirow}

\usepackage{subcaption}
\usepackage{graphicx}

\journal{Nuclear Instruments and Methods: A}

\begin{document}

\begin{frontmatter}

\title{Measurements using a prototype array of plastic scintillator bars for reactor based electron anti-neutrino detection}

\author[a]{P.~K.~Netrakanti\corref{cor1}}
\ead{pawankn@barc.gov.in}
\cortext[cor1]{Corresponding author}
\author[b]{D.~Mulmule}
\author[a]{D.~K.~Mishra}
\author[a]{S.~P.~Behera}
\author[a,c]{R.~Dey}
\author[a]{R.~Sehgal}
\author[d]{S.~K.~Sinha}
\author[a,c]{V.~Jha}
\author[a,c]{and L.~M.~Pant}

\address[a]{Nuclear Physics Division, Bhabha Atomic Research Centre, Trombay, Mumbai - 400085}
\address[b]{Department of Physics, Indian Institute of Technology - Bombay, Mumbai - 400076}
\address[c]{Homi Bhabha National Institute, Anushakti Nagar, Mumbai - 400094}
\address[d]{Research Reactor Services Division, Bhabha Atomic Research Centre, Trombay, Mumbai - 400085}

\begin{abstract}
We report measurement of reactor based electron anti-neutrinos from a prototype array of plastic scintillator bars ( mini-ISMRAN ) located inside Dhruva research reactor hall, BARC. The detector setup took data for 128 days for reactor on (RON) and 51 days for reactor off (ROFF) condition. A detailed analysis procedure is developed to select the anti-neutrino candidate events based on the energy deposition, number of bars hit as well as topological event selection criteria in position and time. Each of these selection criteria are compared with Monte Carlo based simulations and further an embedding technique is used to estimate the efficiencies from a data driven background study. The obtained anti-neutrino like events in RON condition are 218 $\pm$ 50 (stat) $\pm$ 37 (sys) after background subtraction. The obtained results are compared with theoretical estimation which yields 214 $\pm$ 32 (sys) anti-neutrino events for the RON condition.
\end{abstract}

\begin{keyword}
%% keywords here, in the form: keyword \sep keyword

%% PACS codes here, in the form: \PACS code \sep code

%% MSC codes here, in the form: \MSC code \sep code
%% or \MSC[2008] code \sep code (2000 is the default)
  Anti-neutrino \sep Plastic scintillators \sep Reactor monitoring.
\end{keyword}

\end{frontmatter}

%% \linenumbers

%% main text
\section{Introduction}
Measurements of reactor based electron type of anti-neutrinos ( ${\overline{\ensuremath{\nu}}}_{e}$ ) are been carried out by many experimental groups in recent years around the world. Due to a comparatively moderate flux of ${\overline{\ensuremath{\nu}}}_{e}$'s from the reactor and possibility to place detectors, with a volume of few tons, at closer standoff from the reactor core have achieved detailed and precise measurements of ${\overline{\ensuremath{\nu}}}_{e}$ spectral shapes and $\theta_{13}$~\cite{DayaBay,DChooz,RENO,NEOS}. Many of these experiments have tried to understand the so called reactor anti-neutrino anamoly (RAA) from measured data and addressed key information related to the ongoing efforts for the searches of sterile neutrinos at short baselines~\cite{DayaBay,RENOST,PROSPECT}. An observation of excess of ${\overline{\ensuremath{\nu}}}_{e}$ yield in the energy region of 5-6 MeV, when compared with the model predictions, have initiated many theoretical models to take into account the flux uncertainties,  accounting of ${\overline{\ensuremath{\nu}}}_{e}$'s from long lived isotope and effect of isotopic composition of the fuel in a more stringent way~\cite{DB5MeV,RENO5MeV,Huber5MeV}.  
These measurements have also provided a unique tool to monitor reactors in a non-intrusive way and estimate the fuel composition as a function of burn up\cite{SONGS,NUCIFER}. The isotopic fractions for $\mathrm{{}^{235}U}$ and $\mathrm{{}^{239}Pu}$ are estimated from the measurement of an ${\overline{\ensuremath{\nu}}}_{e}$ spectra over a prolonged period of time~\cite{DayaBayFuel}. In recent years, for better understanding of the RAA and sterile neutrino searches, various experiments are now gearing up to perform measurements at closer distances (few meters) from the reactor core. To join the ongoing effort, a plastic scintillator (PS) bar based array, Indian Scintillator Matrix for Reactor Anti-Neutrinos (ISMRAN), will be installed and commissioned in Dhruva reactor hall at $\sim$13 m from the reactor core. The detector setup, will be consisting of an array of plastic scintillator bars arranged in 9$\times$10,  mounted on a movable base structure which will allow us to make the measurements at different distances from the reactor core. The same system will also allow us for the reactor monitoring and determination of isotopic fuel content at different reactor sites. Similar setup using an array of PS bars has been used by the PANDA experiment at a stand off distance of $\sim$25 m from reactor core and have reported promising results~\cite{PANDA}. Also the DANSS collaboration has demonstrated the use of solid plastic scintillator strips for the reactor ${\overline{\ensuremath{\nu}}}_{e}$ detection~\cite{DANSS}.

The ISMRAN detector can observe the active-sterile neutrino mixing sensitivity for $\mathrm{sin^{2}2\theta_{14}}$ $\geq$ 0.064 and $\mathrm{\Delta m^{2}_{41}}$ = 1.0 $\mathrm{eV^{2}}$ at 90$\%$ confidence level for an exposure of 1 ton-year by using neutrinos produced from the DHRUVA reactor with thermal power of 100MWth~\cite{EPJC}. With the combination of an array of plastic scintillator bars at near (~7m) and far (~9m) positions from reactor core, ISMRAN can exclude the active-sterile neutrino mixing in the range 1.4 $\mathrm{eV^{2}}$ $\leq$ $\mathrm{\Delta m^{2}_{41}}$ $\leq$ 4.0 $\mathrm{eV^{2}}$ with present best fit parameters. With the possible region of $\mathrm{sin^{2}2\theta_{14}}$ $\geq$ 0.09 at $\mathrm{\Delta m^{2}_{41}}$ = 1 $\mathrm{eV^{2}}$ for an exposure of 1 ton-year, ISMRAN experiment can observe active-sterile neutrinos oscillations with 95$\%$ confidence level~\cite{Shiba}.
The reactor anti-neutrino spectrum measured using ISMRAN detector setup will be the first of its kind in terms of the fuel composition which pre-dominantly consists of natural uranium. The energy spectral measurement of reactor ${\overline{\ensuremath{\nu}}}_{e}$  from ISMRAN will augment the existing results from similar reactor ${\overline{\ensuremath{\nu}}}_{e}$ experiments which report the 5MeV bump,  mainly attributed to the lack of understanding of the isotopic concentrations and the fuel burnup in the enriched ${}^{235}U$ fuel systems. As compared to the PANDA experiment, the standoff distance for ISMRAN experiment is closer to the reactor core ( d$\sim$13m ), which will help in addressing the sterile neutrino searches using an array of plastic scintllator detector setup.  

In this paper, we present the results from a prototype detector array of PS bars, called mini-ISMRAN, 16$\%$ of the original ISMRAN detector in Dhruva reactor hall. We will also discuss different variables constructed from energy deposits, time and hit position in the PS bars for separating ${\overline{\ensuremath{\nu}}}_{e}$ candidates from the reactor and cosmogenic background. These variables are compared between data and Monte Carlo based simulated events. The Monte Carlo based simulated ${\overline{\ensuremath{\nu}}}_{e}$ events are embedded in the real data to estimate detection efficiencies based on the background characteristics present in the real data. Background subtraction based on two methods will be discussed for obtaining the ${\overline{\ensuremath{\nu}}}_{e}$ candidate events and finally an outlook of the full ISMRAN setup in Dhruva reactor hall is reported.
\section{ISMRAN Detector}
\begin{figure}[h]
\begin{center}
\includegraphics[scale=0.60]{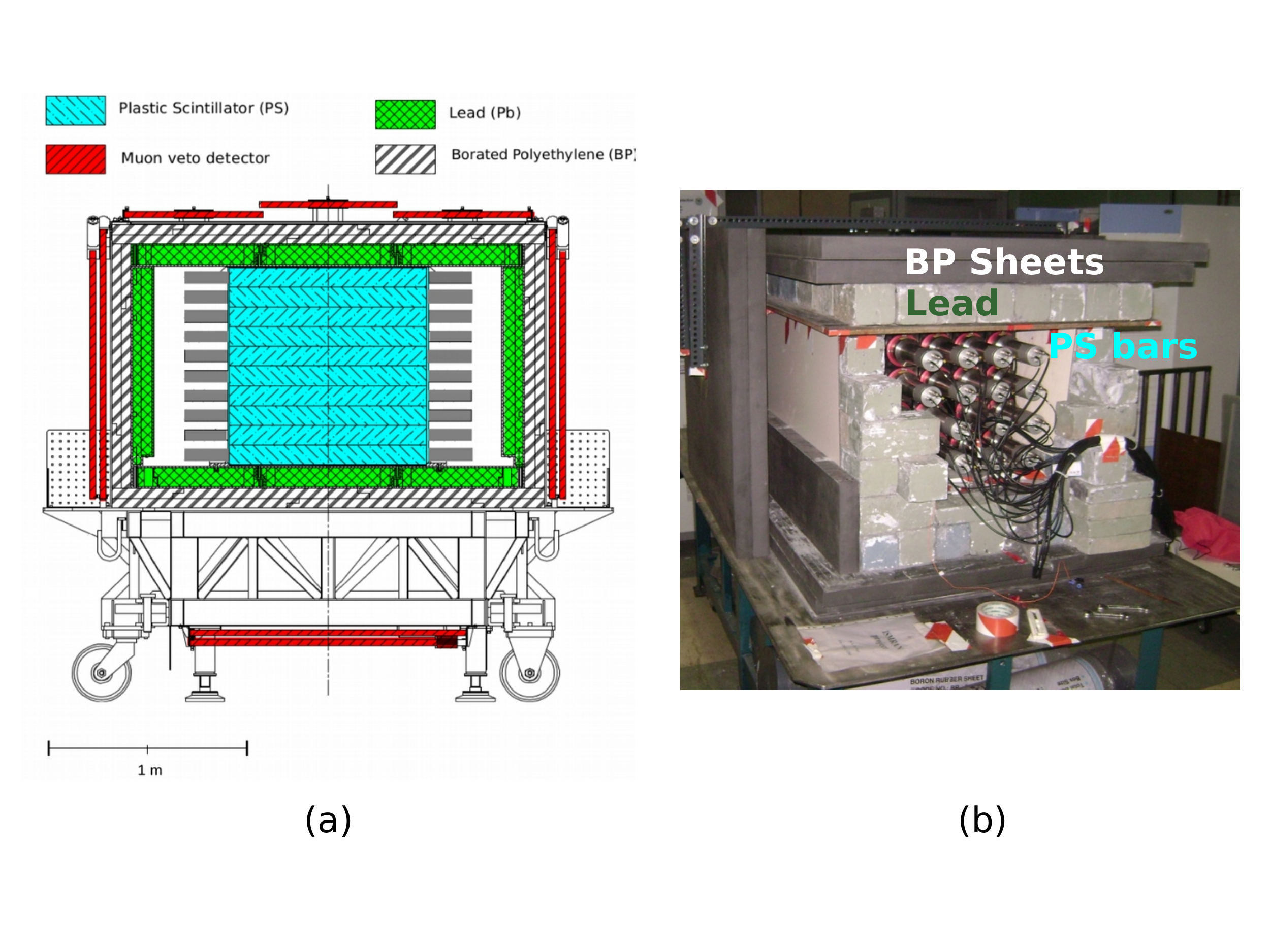}
\caption{Panel (a) : Schematic diagram of full scale ISMRAN experiment consisting of PS bars on a movable base structure with shielding of 10 cm of Pb and 10 cm of BP. Also shown, in red, are the muon veto scintillator detectors for the rejection of comogenic muon induced events.\\ Panel (b) : The mini-ISMRAN prototype setup, in partial shield configuration, at Dhruva reactor hall with 10 cm of Pb and 10 cm of BP shielding.}
\label{fig1}
\end{center}
\end{figure}

A modular array of plastic scintillator bars will form the core detector of the ISMRAN experimental setup~\cite{ISMRAN}. Each PS bar has a dimension of $\mathrm{10 cm \times 10 cm \times 100 cm}$ and are wrapped with Gadolinium oxide ($\mathrm{Gd_{2}O_{3}}$) coated on aluminized mylar foils. The complete detector setup approximately weighing $\sim$1 ton, as shown in Fig.~\ref{fig1} (a), will consist of 90 such bars arranged in a 9$\times$10 matrix. Each PS bar has a composition similar to EJ200~\cite{eljen} and is directly coupled to a 3 inch Photomultiplier tubes (PMT) at both ends for signal readout of the triggered events. A digital data acquisition system consisting of CAEN VME based waveform digitizers (V1730) will be used to filter the triggered events from each PS bar independently. The digitizer channels are capable of processing the anode pulses from PMTs at 500 MS/s with on board implementation of trigger generation, threshold selection, constant fraction discrimination as well as integration of the accumulated charge, in units of ADC, for individual PMTs. A unique timestamp is allocated to each triggered event, which is used for offline building of correlated events. The use of the digitizers allows us to collect data practically with zero dead time with the data taking rates reaching as high as few kHz in each PS bar. 
The PS bars are enclosed in a shielding of 10 cm thick of Lead (Pb) and 10 cm thick of borated polyethylene (BP), with 30$\%$ boron concentration, in a layered manner. To get the optimal configuration of layered shielding, the Pb and BP shielding thicknesses are optimized along with the maximum weight handling capacity of the reactor floor. From simulation studies, a rejection of $\sim$98$\%$ of $\gamma$-ray background events and $\sim$90$\%$ of fast neutron events below 10 MeV, is achieved with the layered shielding~\cite{ISMRAN}.  An active muon veto plastic scintillator detectors, covering ISMRAN from all sides, will be installed for the rejection of cosmic muon and muon induced background events. The entire setup, including shielding, will be housed on a movable base structure, also shown in Fig.~\ref{fig1} (a), approximately at a distance of $\sim$13 m from the Dhruva reactor core. The ${\overline{\ensuremath{\nu}}}_{e}$ spectral measurements at various distances inside reactor hall will allow us to address the physics of sterile neutrino searches~\cite{Shiba}.
Dhruva research reactor, a natural uranium fuel based thermal reactor ($\mathrm{100 MW_{th}}$) situated in Bhabha Atomic Research Center, Trombay, is primarily used for the research involving the neutrons produced from the fission reactions. It also helps in catering a wide variety of radio-isotopes for the medical applications across the country. More details about the Dhruva reactor can be found in following Ref.~\cite{DHRUVA}.

A prototype detector, mini-ISMRAN consisting of 16 PS bars, has been set up at the allocated site of the main ISMRAN experiment in Dhruva reactor hall for the measurements of background rates in RON and ROFF conditions~\cite{ISMRAN}. In Fig.~\ref{fig1} (b) shown is the prototype mini-ISMRAN setup in Dhruva reactor hall. The arrangement of shielding structure is shown in a cross-sectional view along with the PS bars. For the final data taking, the mini-ISMRAN is completely enclosed in the shielding from all around the PS bars. The mini-ISMRAN took data approximately a year long period in year 2018 and various measurements are performed for the study of stability of the detector system, development of the data acquisition system as well as understanding of the effect of shielding around the detector system. Due to an above ground setup and in close vicinity of the reactor core, the mini-ISMRAN faced the harsh backgrounds from the reactor surroundings. These backgrounds mainly consisted of fast neutrons and $\gamma$-rays predominantly coming from the thermal neutron captures on the surrounding structural materials. A shielding of 10 cm of Pb and 10 cm of BP is erected around mini-ISMRAN all over with an extra 5 cm of high density polyethylene, installed later and not shown in Fig.~\ref{fig1} (b), from the outer side. 
With continuous running mode of mini-ISMRAN, the recorded data is analyzed for the reconstruction of the ${\overline{\ensuremath{\nu}}}_{e}$ candidate events from a dataset consisting of 128 days of RON and 51 days of ROFF data.
The reactor based ${\overline{\ensuremath{\nu}}}_{e}$ events in mini-ISMRAN PS bars are detected using the inverse beta decay (IBD) process. The IBD interaction of ${\overline{\ensuremath{\nu}}}_{e}$ with protons present in PS bar generates a positron and a neutron as daughter particles, shown in Eq.~\ref{eq:ibd}.
\begin{equation}\label{eq:ibd}
\mathrm{ \overline \nuup_{e} + p \rightarrow e^{+} + n}.
\end{equation}
The positron being lighter in mass as compared to neutron shares the maximum amount of the ${\overline{\ensuremath{\nu}}}_{e}$'s energy, leaving a few keVs of energy for the neutron. The positron energy loss in the PS bars and subsequent annihilation while coming to rest, emitting two $\gamma$-rays of 0.511 MeV each, comprise the prompt event. On the other hand, the neutron loses its energy through multiple scattering and eventually thermalize in the scintillator volume and get captured on either on Gadolinium (Gd) foils or on Hydrogen (H) of the scintillator medium. The above mentioned IBD process has an energy threshold of 1.806 MeV, hence only reactor ${\overline{\ensuremath{\nu}}}_{e}$ above this threshold energy can be detected in the ISMRAN setup.
\begin{equation}\label{eq:Gd1}
  \hspace{-0.2in}
\mathrm{ n + {}^{155}Gd \rightarrow {}^{156}Gd^{*} \rightarrow \gamma 's,   \quad \sum E_{\gamma} = 8.5~MeV}, \quad \sigmaup_{\mathrm{n-capture}} = \mathrm{61000~b},
\end{equation}
\begin{equation}\label{eq:Gd2}
  \hspace{-0.2in}
\mathrm{ n + {}^{157}Gd \rightarrow {}^{158}Gd^{*} \rightarrow \gamma 's,  \quad \sum E_{\gamma} = 7.9~MeV}, \quad \sigmaup_{\mathrm{n-capture}} = \mathrm{254000~b}.
\end{equation}
Due to larger thermal neutron capture cross-section for Gd, as shown in Eqs.~\ref{eq:Gd1} and ~\ref{eq:Gd2}, from simulations it has been observed that $\sim$75$\%$ of the events are captured on Gd and rest $\sim$25$\%$ events are captured on H in ISMRAN detector geometry~\cite{ISMRAN}. The thermal neutron capture on the Gd or H defines the delayed event. The emission of cascade $\gamma$-rays, from the de-excitation of Gd nuclei, with a sum energy of up to $\sim$8 MeV or the neutron capture on H emitting a mono-energetic $\gamma$-ray of 2.2 MeV will be the delayed event signature. By looking at the sum energy depositions for the prompt positron and delayed neutron capture events, we can reconstruct the ${\overline{\ensuremath{\nu}}}_{e}$ candidate events in ISMRAN. Also, the characteristic time difference ($\mathrm{\tau}$) between the prompt and the delayed event signatures, of the order of $\sim$68 $\mathrm{\mu}$s for the ISMRAN detector geometry obtained from simulations, helps in discriminating the ${\overline{\ensuremath{\nu}}}_{e}$ from other correlated and uncorrelated background events~\cite{ISMRAN}.
Using the prescription from Ref.~\cite{VVER}, an estimate of $\sim$60 ${\overline{\ensuremath{\nu}}}_{e}$ events per day is predicted with a detection efficiency of 16$\%$~\cite{ISMRAN} for ISMRAN detector with Dhruva reactor operated at 100 $\mathrm{MW_{th}}$. The estimation in ${\overline{\ensuremath{\nu}}}_{e}$ event rate is for the point source as opposed to the elongated core of Dhruva reactor. 

For understanding the response of the PS bars, estimation of detection efficiencies, shielding effects and study of various backgrounds,  Monte Carlo based simulated events are passed through GEANT4 (version 4.10.04) package~\cite{GEANT4}. The standard electromagnetic processes are used for the response of $\gamma$-rays, electrons and positrons. The high precision QGSP\_BIC\_HP physics processes are used for the low energy neutron interactions. For cascade $\gamma$-ray emission from the thermal neutron capture on Gd nucleus, GEANT4 uses photon evaporation model which conserves the final sum energy of the captured event, but does a poor modeling of the individual $\gamma$-ray energy distribution on an event by event basis~\cite{ANRIGD}. To overcome this problem in our simulation studies, we have used the DICEBOX package for simulating the cascade $\gamma$-rays from the thermal neutron capture on Gd nucleus~\cite{DICEBOX}.  For a reactor based ${\overline{\ensuremath{\nu}}}_{e}$ spectra as an input to GEANT4 package, the events are generated using parameterization from Ref.~\cite{Huber,Mueller}, the calculation of cross sections from Ref.~\cite{Mention,Vogel} and the fission fractions for different isotopes are taken from Ref.~\cite{Zhan}. The positrons and neutrons produced are assigned their energy and directions kinematically from the ${\overline{\ensuremath{\nu}}}_{e}$ energy and then propagated through GEANT4 for recording of their interactions and energy depositions in the PS bars. The PS bars are accounted for the detector response. This includes the smearing of true energy in each individual PS bar with the energy resolution obtained by comparison of the response of PS bar between data and MC events for the $\mathrm{{}^{137}Cs}$, $\mathrm{{}^{22}Na}$ and $\gamma$ from an AmBe source. The timing and position resolutions are also incorporated using the measured data with Cs source in each PS bar. In future studies for the full scale ISMRAN setup, the light propagation in the PS bars will also be incorporated in the simulating the IBD event signatures. All the results presented from the Monte Carlo based simulations are termed as MC now onwards in the paper.
\begin{figure}[h]
\begin{center}
\includegraphics[scale=0.80]{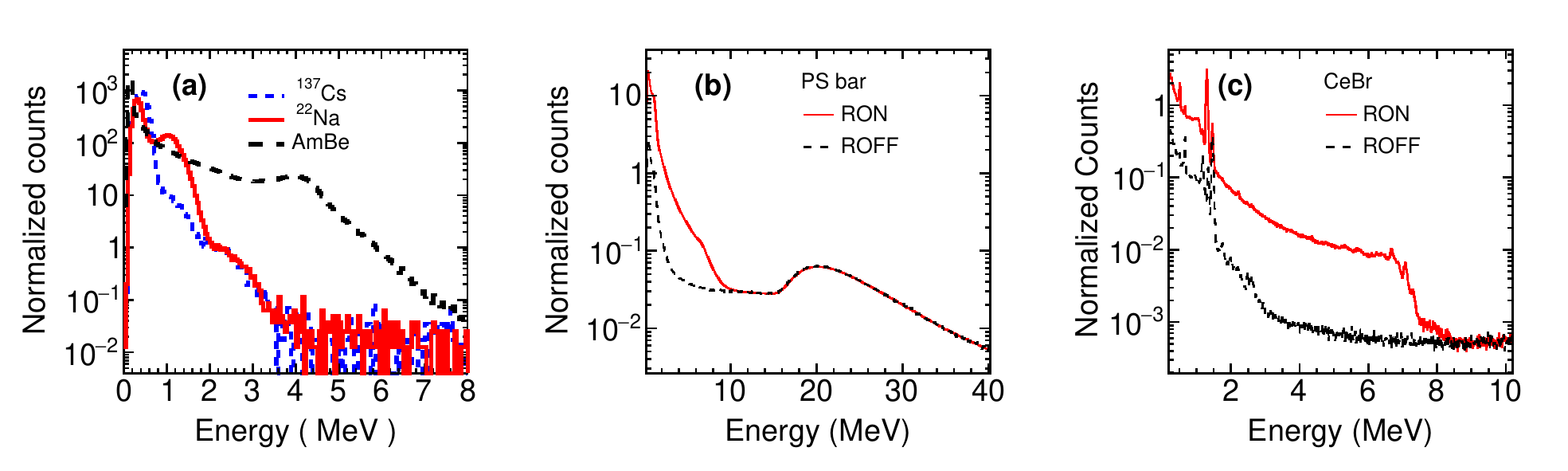}
\caption{Energy distribution and response for (a) various radioactive sources and (b) RON and ROFF condition for a PS bar. Panel (c) shows the $\gamma$-ray energy distribution from a CeBr detector in RON and ROFF condition.}
\label{fig2}
\end{center}
\end{figure}
The energy calibration of all the PS bars are performed using the Compton edges from the standard $\gamma$-ray sources consisting of ${}^{137}$Cs, ${}^{60}$Co and ${}^{22}$Na. For higher energy calibration point we have used the 4.4 MeV $\gamma$-ray from the AmBe source. The energy response for the ${}^{137}$Cs source in one single PS bar is compared between data and MC simulation and an energy resolution of $\sim$18$\%$ /$\mathrm{\sqrt{E}}$ is obtained~\cite{ISMRAN}. Figure~\ref{fig2} (a) and (b) show the energy response of the PS bar for the standard radioactive source in a non-reactor environment and spectra from reactor hall in RON and ROFF conditions, respectively. From Figure~\ref{fig2} (b), it can be seen that in ROFF condition, the dominant background up to 3 MeV is from the natural radioactivity from ${}^{40}$K and ${}^{208}$Tl and above that the spectrum is dominated by the cosmogenic background from muons. However, in RON condition up to 8 MeV, the background is not only dominant from the natural radioactivity from the ${}^{40}$K and ${}^{208}$Tl but also from the $\gamma$-rays emitted from either fissioning isotope daughter nuclei or from thermal neutron captures on the surrounding structural material inside the reactor hall. The cosmic muon energy deposition ($\mathrm{E^{\mu}_{dep}}$) can be seen around $\sim$20 MeV in both RON and ROFF condition and the scaling of muon event rates between RON and ROFF indicates the independence of operating power of the reactor and cosmogenic background above 10 MeV energy region in the reactor hall.
Figure~\ref{fig2} (c) shows the $\gamma$-ray spectrum from a Cerium Bromide (CeBr) detector, which has a better energy resolution $\sim$4$\%$ at 0.662 MeV, for RON and ROFF condition. It can be seen that in ROFF condition the emission of $\gamma$-rays from ${}^{137}$Cs, ${}^{60}$Co along with the natural radioactivity from ${}^{40}$K and ${}^{208}$Tl below 3 MeV. In RON condition, from Fig.~\ref{fig2} (c), the dominant $\gamma$-ray background comes from ${}^{41}$Ar in the energy region below 3 MeV. In the region of 6-8 MeV, the $\gamma$-rays from a thermal neutron capture on the near by iron and copper structures is the dominant source of the RON background which is also reported in the Ref.~\cite{ReactorBkg}. The composition of the these structural elements are mostly similar in a typical reactor environment.
%very precise $\gamma$-ray spectrum up to 8 MeV, with an excellent energy resolution HPGe detector, measured by PROSPECT collaboration shows the relative $\gamma$-ray background yields from different structural elements in a RON condition in a HFIR~\cite{ReactorBkg}. 
\begin{figure}[h]
\begin{center}
\includegraphics[scale=0.65]{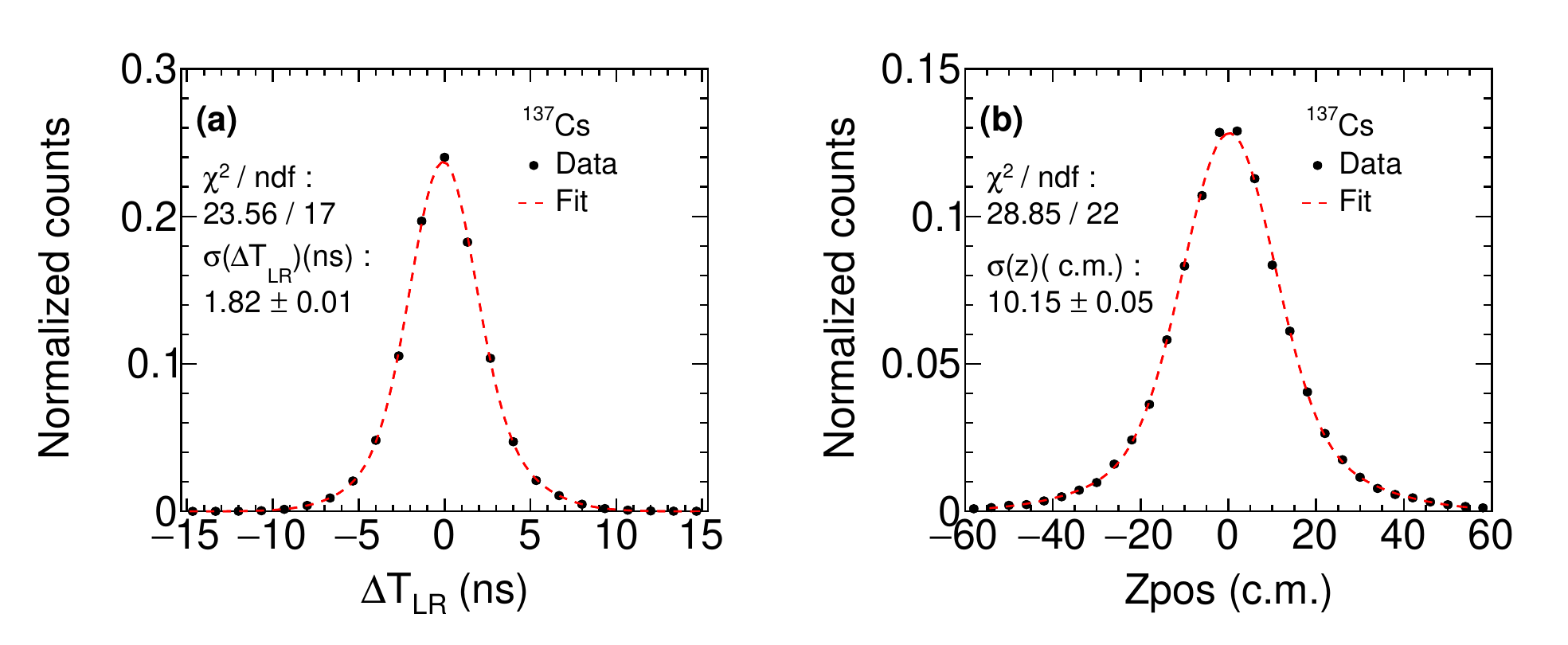}
\caption{(a) Timing difference $\mathrm{\Delta T_{LR}}$ and (b) position ($\mathrm{Zpos}$) distribution in a PS bar measured with ${}^{137}$Cs source kept at the center of the PS bar. Also shown, in red, are a double Gaussian fit to the distribution to extract the timing and position resolutions.}
\label{fig3}
\end{center}
\end{figure}

Using the timestamp information from both the PMTs coupled at the end of each PS bar, we can reconstruct the timing difference ($\mathrm{\Delta T_{LR}}$) and parameterize an approximate position ($\mathrm{Zpos}$) along the length of the PS bar by taking measurements at different positions with a radioactive source~\cite{ISMRAN}. The center of the PS bar is taken as 0 cm and each end of PS bar spans -50 cm and +50 cm in length. Figure~\ref{fig3} (a) and (b) show the ($\mathrm{\Delta T_{LR}}$) and $\mathrm{Zpos}$ distributions for a PS bar when ${}^{137}$Cs source is kept at the center (0 cm) of the PS bar, respectively. Both the distributions are fitted with a double Gaussian function to extract the timing and position resolution for the PS bars. The timing resolution of $\sim$2 ns leads to a position resolution of $\sim$10 cm in a PS bar. The position resolution along a single PS bar is almost constant up to $\pm$ 30 cm and varies by $\sim$6$\%$ there after towards the end of the PS bar. This is mainly due to the nonlinear effects caused by the light attenuation which starts to show up at the edge of the PS bar. 
Determination of the timing and position resolutions in a PS bar are mainly affected by the inability for a proper collimation of the radioactive source along the length of the PS bar. These resolutions represent an upper limit from the measured data using radioactive source only. The timing and position resolutions are mostly similar in all the bars and are independent of incident $\gamma$-ray energy. Due to a segmented geometry of the ISMRAN detector, we can exploit the timing and position information for the localization of an event in the ISMRAN geometry to differentiate between an ${\overline{\ensuremath{\nu}}}_{e}$ and a background event from fast neutron~\cite{MLP,Roni}.

In the reactor hall, the stability for all the PS bars are checked using the $\mathrm{E^{\mu}_{dep}}$ from cosmic muons during the data taking mode in RON and ROFF conditions.  
\begin{figure}[h]
\begin{center}
\includegraphics[scale=0.65]{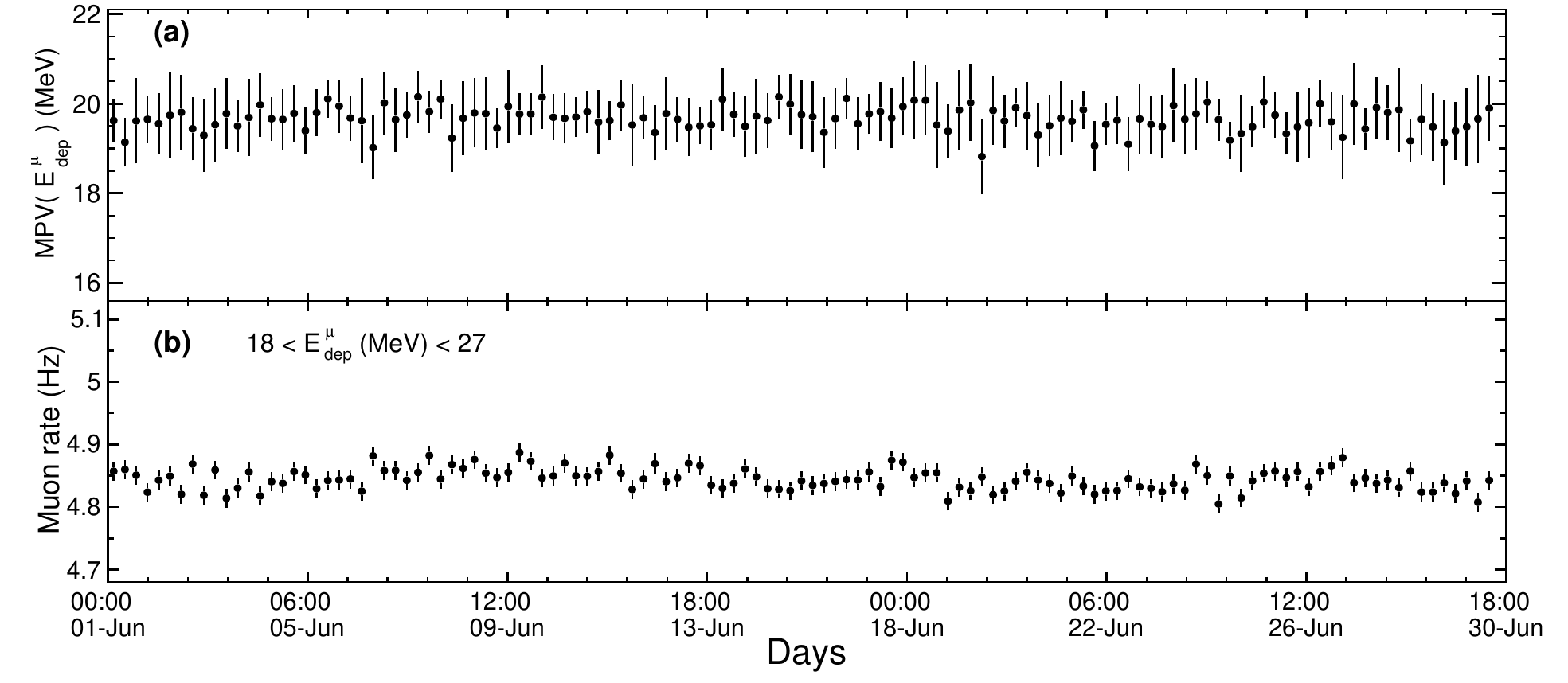}
\caption{Panel (a) : Stability of a single PS bar, in reactor on condition, in reactor hall as a function of time using the cosmic muon energy deposition. Panel (b) : Cosmic muon rate in a single PS bar within the energy range 18 MeV $<\mathrm{E^{\mu}_{dep}}<$ 27 MeV as a function of time in reactor hall.}
\label{fig4}
\end{center}
\end{figure}
The $\mathrm{E^{\mu}_{dep}}$ in each PS bar, as seen from Fig.~\ref{fig2} (b), is fitted using a convoluted Landau with a Gaussian function. Figure~\ref{fig4} (a) shows the most probable value of $\mathrm{E^{\mu}_{dep}}$ in the PS bar as a function of time for the month of June. The errors shown are statistical and fit errors combined in quadrature. The stability and performance of the detector throughout the data taking period is reasonable. This  stability check is done for all the PS bars during the entire running period in reactor hall. Figure~\ref{fig4} (b) shows the average rate of cosmic muons reconstructed from the PS bar in the energy range of 18 MeV $<\mathrm{E^{\mu}_{dep}}<$ 27 MeV. These are the efficiency uncorrected rates and are good enough for the stability check of the PS bars in reactor hall. Out of total 16 PS bars, two PS bars were showing gain drifts as a function of time and are hence masked from the data analysis as well as in the MC event generation. The gain drift in these two PS bars was found to be caused by the faulty potentiometer between the high voltage source and the PMT base. 

\section{Prompt and delayed events}
As discussed previously, the prompt event comprises of the energy deposition from the positron and the annihilation $\gamma$-rays. Similarly, the delayed event signature consists of the energy deposited by the cascade $\gamma$-rays from the n-Gd capture process. The sum energy variable is constructed by grouping the individual PS bars, with energy deposition ( $\mathrm{E_{bar}}$ ) between 0.2 and 5.5 MeV, in a 20 ns time window allowing us to construct the prompt and delayed event signatures~\cite{ISMRAN}. The lower energy threshold is chosen for the uniformity of the reconstructed energy for all the PS bars. The higher energy threshold of 5.5 MeV is chosen to remove the $\gamma$-ray background from neutron capture on surrounding material in the reactor hall, which is the dominant source of background above 6.0 MeV. The number of bars hit ( $\mathrm{N_{bars}}$) satisfying above condition is another variable for discriminating the prompt and delayed events from the background events. 

From the data, we sort PS bars in ascending order of their timestamps and then build event according to closeness in time. The event building time window is chosen as as 20 ns, beyond which the inter-bar timing difference from data shows the uniform distribution in time representing the randomness of the PS bar events. Once an event is built, according to the sum energy deposition ($\mathrm{E_{sum}}$) and number of PS bars hit ($\mathrm{N_{bars}}$), it is assigned either as a prompt event or as a delayed event candidate. For prompt event candidate, the sum energy, $\mathrm{E^{p}_{sum}}$, is required to be in the range of 2.8 to 6.0 MeV and the number of bars hit, $\mathrm{N^{p}_{bars}}$ should be either 2 or 3.
The sum energy criteria are mainly driven by the smaller volume of the mini-ISMRAN detector. In the full scale ISMRAN data analysis, the sum energy selection criteria for the prompt events will be relaxed since there would be better localization of the reconstructed prompt event candidates from IBD event in the full detector volume. From simulation studies, the impact on the spectral shape of the anti-neutrino spectra is estimated to be $\sim$10$\%$ because of the energy selection criteria with current reconstruction scheme.
Similarly, for the delayed event candidate selection, a sum energy ($\mathrm{E^{d}_{sum}}$) between 3.0 to 6.0 MeV and the number of bars hit ($\mathrm{N^{d}_{bars}}$) should be greater than equal to 4 or less than equal to 6. The lower and higher energy thresholds on the $\mathrm{E^{p}_{sum}}$ and $\mathrm{E^{d}_{sum}}$ for prompt and delayed event candidates are chosen to reduce the contamination from the reactor related background as well as contamination from the natural radioactive background. Though these stringent criteria reduces the detection efficiency of the prompt and delayed event candidates, but in turn allows us to suppress the reactor related backgrounds considerably. Apart from choosing prompt and delayed event candidates based on their $\mathrm{E_{sum}}$ and $\mathrm{N_{bars}}$, a rejection criteria is applied to remove prompt or delayed events which have occurred within a time window of 250 $\mathrm{\mu}$s from a cosmic muon event in any of the PS bar. This selection criteria would allow us to avoid the muon decay events in the mini-ISMRAN detector. 

All those prompt-delayed candidate pairs are rejected if there lies another prompt or delayed event candidate, in time, between the selected prompt and delayed event candidate. Similarly all those prompt and delayed candidate events are rejected when a subsequent prompt or delayed event is found within a window of 500 $\mathrm{\mu}$s of the original selected prompt or delayed event. These are identified as fake prompt or delayed events and their rates are dependent on the reactor power. 
To understand the effect of power dependence on the background, we use a technique called embedding to estimate the prompt, delayed and ${\overline{\ensuremath{\nu}}}_{e}$ event detection efficiencies. Embedding of MC events in real data is commonly used in high energy heavy-ion physics experiments to estimate efficiencies in a data driven background scenario~\cite{STAR}. This technique involves the embedding of the prompt positron and delayed neutron event, generated from the reactor ${\overline{\ensuremath{\nu}}}_{e}$ spectra and decayed kinematically, in the real data at the level of PS bars. Since, the MC generated events lack the timestamp information, the PS bars forming prompt and delayed events are assigned timestamps in a manner such that they match the timestamp spread of the individual PS bar events obtained from the real data. The energy, timing  and position resolutions are propagated to the PS bars in the embedded event. The characteristic time difference ($\mathrm{\tau}$) in the MC generated events is preserved between the prompt and the delayed pair as assigned by the GEANT4. A total of 25 events per second are embedded in real data. We have changed the number of embedded events from 10 to 100 and do not see any significant change in the reconstruction efficiencies. Embedding of MC generated events in real data allows us to estimate the rejection of true prompt and delayed event candidates due to the overlap of the reactor related and cosmogenic background rate. A loss of $\sim$4$\%$ in efficiency is observed due to the rejection criteria chosen from cosmogenic muon background and $\sim$9$\%$ due to the fake prompt and delayed pairs reconstructed in the data.
\begin{figure}[h]
\begin{center}
\includegraphics[scale=0.65]{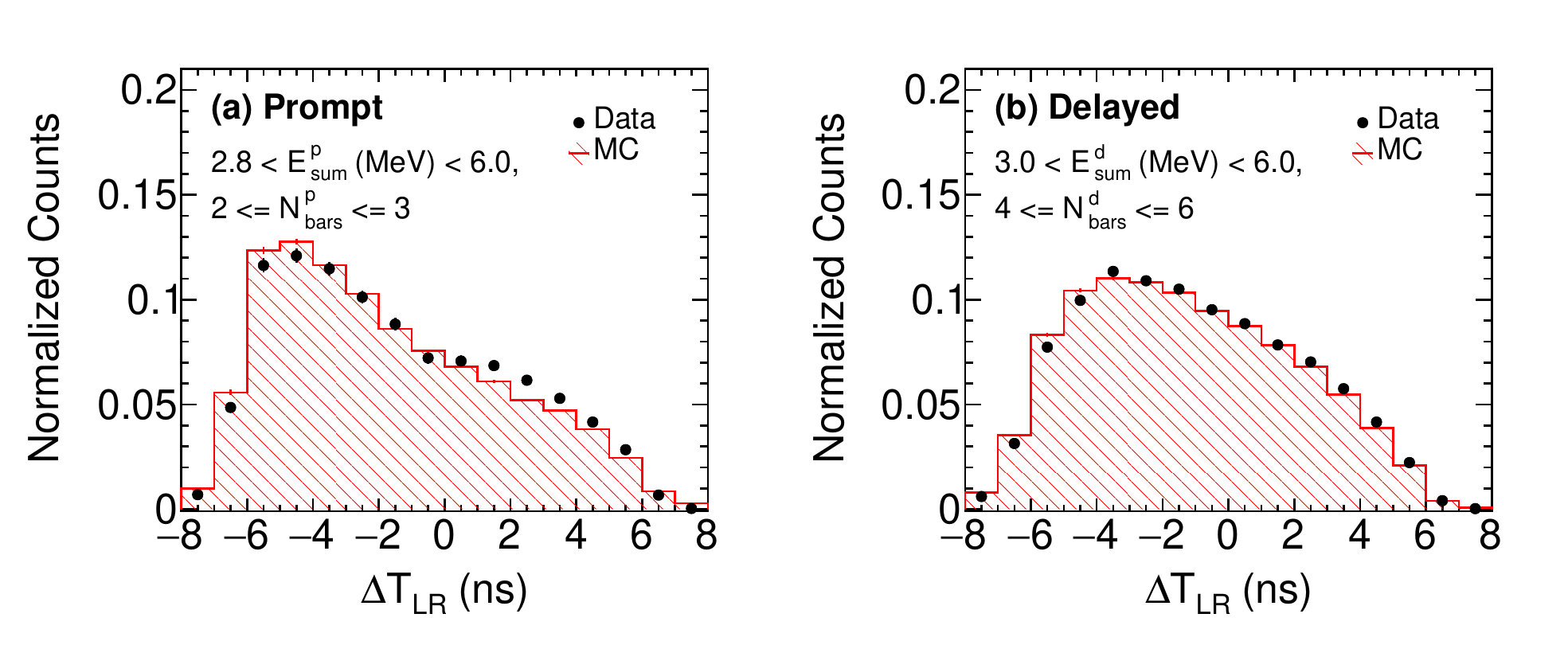}
\caption{Comparison between data (solid) and MC (shaded) for $\mathrm{\Delta T_{LR}}$, intra-bar timing difference between two PMT's of the PS bars in (a) prompt and (b) delayed events.}
\label{fig5}
\end{center}
\end{figure}
Apart from the sum energy and number of bars hit selection variables, the spread of PS bars in prompt and delayed event candidates in time and position are also included in the selection variables. The maximum energy deposition in a single PS bar, ratio of maximum energy deposition to the sum energy and ratio of energy depositions among the PS bars for prompt and delayed events can be used to filter out the signal from background events. Similar analysis level event selection criteria are used by PANDA collaboration~\cite{PANDA}. A selection criteria on the above mentioned variables, may lead to better signal to background discrimination, but needs to be compared in detail with the MC simulation results before any event selection are made from such a cut based analysis. We define ``prompt'' and ``delayed'' now onwards for the prompt and delayed event candidates otherwise mentioned explicitly. All the plots for the variables are shown for the RON condition only. Similar plots are obtained for the ROFF condition and the efficiency numbers are comparable between two conditions.
Figure~\ref{fig5} (a) and (b) show the intra-bar timing difference between the PMTs ($\mathrm{\Delta T_{LR}}$) of the PS bars for the prompt and delayed event candidates, respectively. The comparison are done between the data and MC generated events. There is an asymmetry in the $\mathrm{\Delta T_{LR}}$ distribution, which is attributed to the directional background contamination due to the presence of a neutron guide tube in the vicinity of the mini-ISMRAN experiment. This asymmetry is propagated in the MC simulation events and the MC results shows a reasonable agreement with data. A more robust shielding arrangement is been prepared for the guide tube for the future measurements with the full scale ISMRAN detector. We consider Z direction along the length of the PS bar for all the further studies in the paper. Figure~\ref{fig6} (a) and (b) show the reconstructed Z position ($\mathrm{Zpos}$) for the PS bars, parameterized from the $\mathrm{\Delta T_{LR}}$, for prompt and delayed events, respectively. The timing resolution has been translated into $\mathrm{Zpos}$ resolution and is reflected in the MC events. All the PS bars which are hit within $\pm$ 40 cm along $\mathrm{Zpos}$  are selected for reconstruction of sum energy for prompt or delayed events.
\begin{figure}
\begin{center}
\includegraphics[scale=0.65]{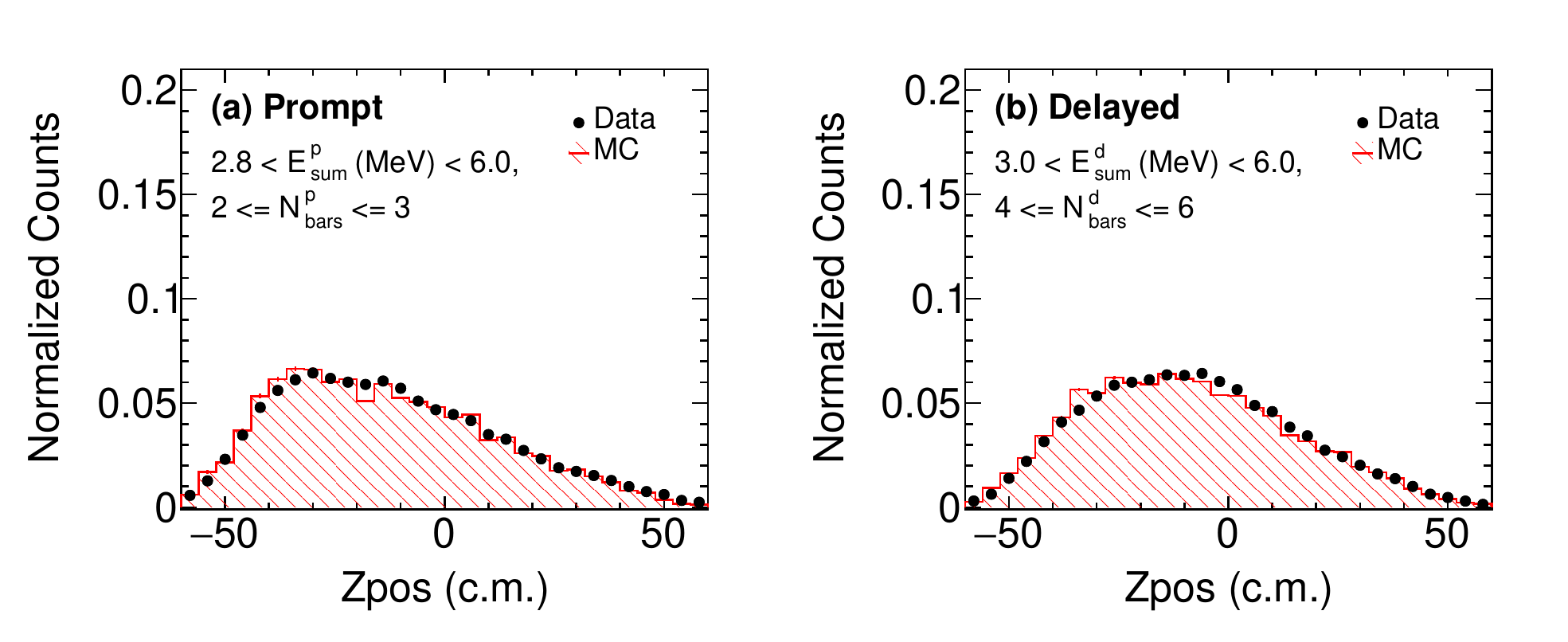}
\caption{Comparison between data (solid) and MC (shaded) parameterized Z position ($\mathrm{Zpos}$) from the $\mathrm{\Delta T_{LR}}$ of PS bars in (a) prompt and (b) delayed events.}
\label{fig6}
\end{center}
\end{figure}

\begin{figure}[h]
\begin{center}
\includegraphics[scale=0.65]{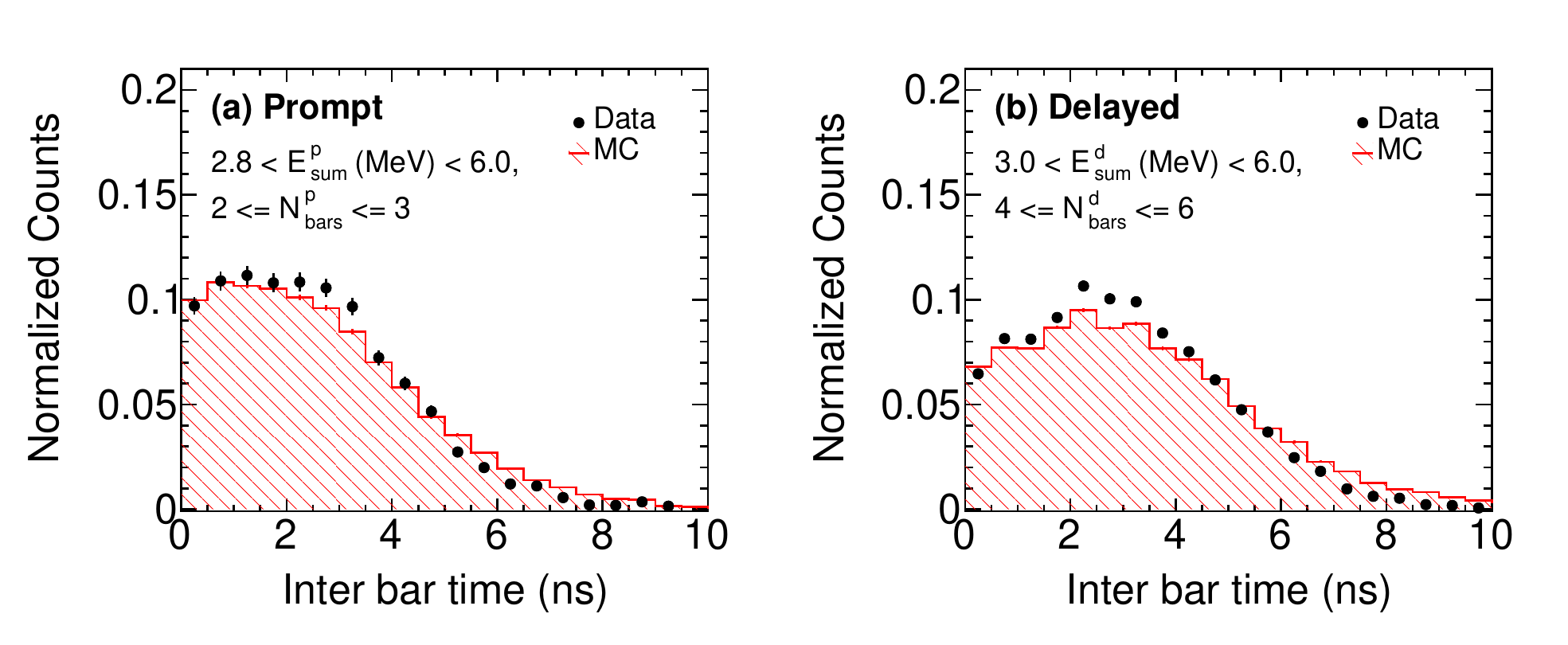}
\caption{Comparison between data (solid) and MC (shaded) for inter-bar time spread of PS bars in (a) prompt and (b) delayed events.}
\label{fig7}
\end{center}
\end{figure}
\begin{figure}[h]
\begin{center}
\includegraphics[scale=0.65]{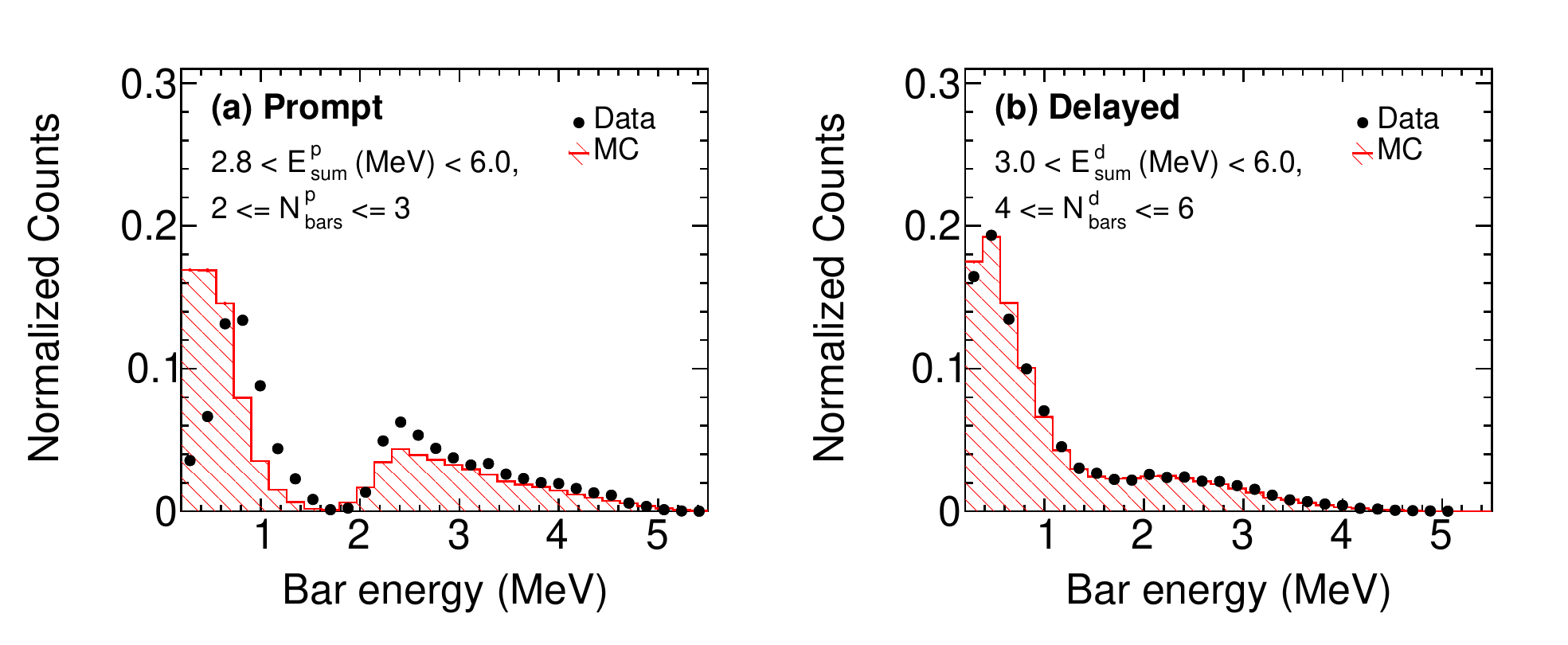}
\caption{Comparison between data (solid) and MC (shaded) for PS bar energy in (a) prompt and (b) delayed events.}
\label{fig8}
\end{center}
\end{figure}

The time spread among the PS bars in prompt and delayed event is also selected to filter out the events of interest. Figure~\ref{fig7} (a) and (b) show the inter-bar time spread in prompt and delayed events, respectively. It can be seen that these events are tightly correlated and a selection cut of 10ns would reduce any uncorrelated accidental background.
Since the prompt and delayed candidates are identified on the basis of $\mathrm{E_{sum}}$, it is important to compare the individual PS bar energy distribution between data and MC events for prompt and delayed candidates. Figure~\ref{fig8} (a) and (b) show the individual PS bar energy distribution which forms the the prompt and delayed events after applying the $\mathrm{E_{sum}}$ and $\mathrm{N_{bars}}$ selection criteria, respectively. It can be seen that there is a disagreement between data and MC events below 1.5 MeV in prompt events and can be understood in terms of low energy background mainly from the reactor or natural radioactivity. 
These natural or reactor related background events can also span 2 or 3 bars, satisfying the $\mathrm{E^{ p}_{sum}}$ condition in prompt candidate events. Though the embedding of the MC generated IBD events in the real data background provides an estimate of the reactor related background contaminating the prompt event candidate sum energy reconstruction, a detailed study of these events in the lower energy region for the PS bars needs to be performed in future analysis with the full scale ISMRAN setup. Due to the limited statistics in the ROFF condition, nothing conclusive can be addressed about this feature in the lower energy of the PS bars in the current results. 
Above 2 MeV, the comparison between the data and MC events for prompt events is reasonable. The dip in the individual PS bar energy distribution for the prompt events at $\sim$ 1.8 MeV is mainly due to the IBD threshold and the selection criteria depending on the $\mathrm{E^{ p}_{sum}}$ and $\mathrm{N^{ p}_{bars}}$ for prompt events.
For the neutron capture delayed events the agreement is much better between data and MC events, as already a tighter selection criteria is applied in  event selection, which would in turn reduce the contribution from the reactor or natural radioactive background $\gamma$-rays.
\begin{figure}[h]
\begin{center}
\includegraphics[scale=0.65]{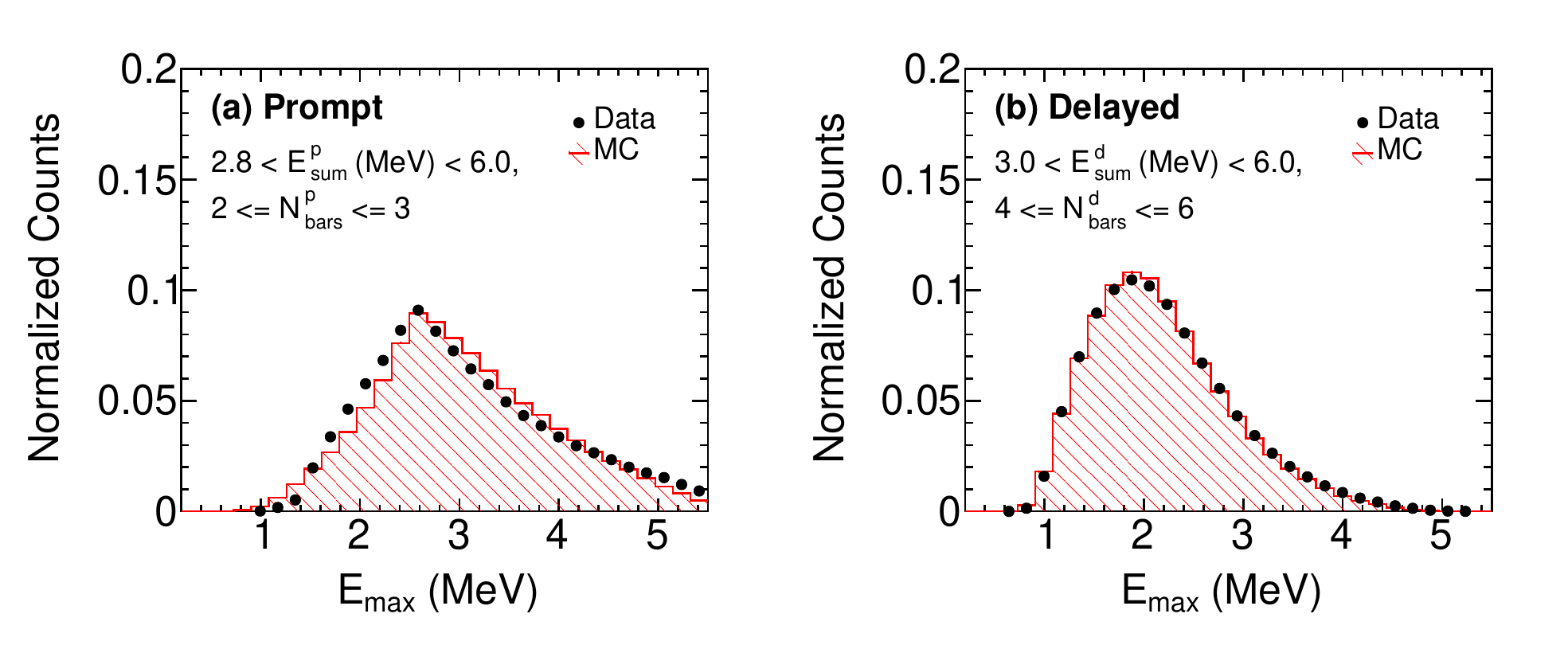}
\caption{Comparison between data (solid) and MC (shaded) for maximum energy deposited ($\mathrm{E_{max}}$) in a PS bar in (a) prompt and (b) delayed events.}
\label{fig9}
\end{center}
\end{figure}
Since the prompt or delayed event candidates are reconstructed having a criteria of at least multiple PS bars are hit which satisfies the respective sum energy selection. We sort the hit PS bars in each candidate event, according to the decreasing order of their individual energy depositions. The PS bar which has maximum energy deposition, is denoted as $E_{max}$, out of all the PS bars used in the reconstruction of the prompt or delayed candidate event. Similarly $E_{1}$ represents the the PS bar which is next, in terms of energy deposition with respect to the $E_{max}$, in the reconstructed candidate event. 
To further understand the individual PS bar energy profiles in prompt and delayed event, we compare the PS bar with maximum energy deposited ($\mathrm{E_{max}}$) between data and MC events. Figure~\ref{fig9} (a) and (b) show the $\mathrm{E_{max}}$ distribution for prompt and delayed events, respectively. There seems to be reasonable agreement between data and MC events for the prompt events in the higher energy regions and the discrepancy observed in Fig.~\ref{fig8} (a) is mainly due to the reactor related or natural background below 3 MeV. The agreement between data and MC for the delayed event is reasonable and hence benchmarks the n-Gd capture model implemented using DICEBOX in GEANT4 for the mini-ISMRAN.
\begin{figure}[h]
\begin{center}
\includegraphics[scale=0.65]{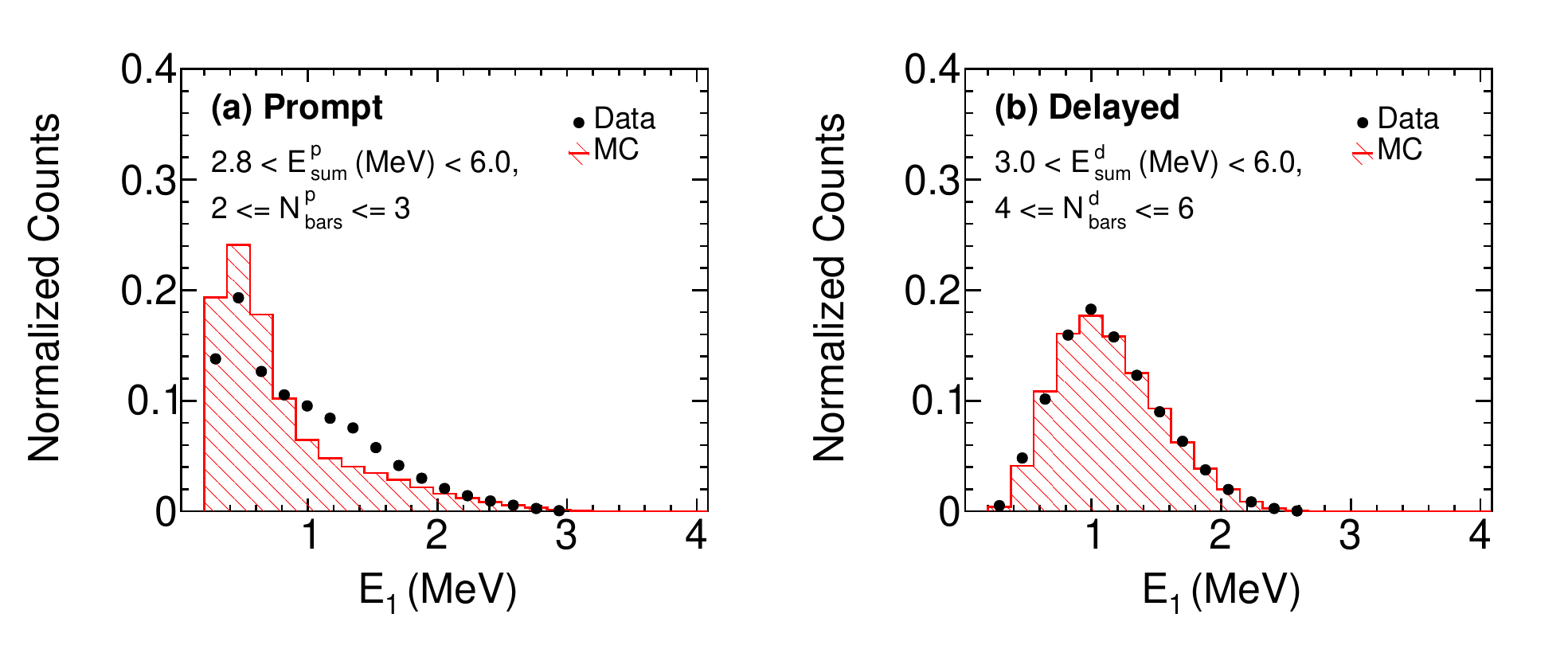}
\caption{Comparison between data (solid) and MC (shaded) for $\mathrm{E_{1}}$ in a PS bar for (a) prompt and (b) delayed events.}
\label{fig10}
\end{center}
\end{figure}
Similarly, we also compare energy deposition ($\mathrm{E_{1}}$), the following PS bar next to $\mathrm{E_{max}}$ in terms of the energy deposition, as shown in Fig.~\ref{fig10} (a) and (b) for prompt and delayed event, respectively. As can be seen from Fig.~\ref{fig10} (a) there are excess of events in data as compared to MC between 0.8 and 2.0 MeV region, which is mainly the region of contamination from $\gamma$-rays produced from the natural radioactive ${}^{40}$K and ${}^{208}$Tl and reactor related background. Again for the delayed events, shown in Fig.~\ref{fig10} (b), the data and MC events comparison seems to be in good agreement.
More differential comparison of the PS bar energy profiles is done using the ratio of $\mathrm{E_{max} / E_{sum}}$ and $\mathrm{E_{1} / E_{max}}$ for the prompt and delayed events. Figure~\ref{fig11} (a) and (b) show the $\mathrm{E_{max} / E_{sum}}$ distribution for the prompt and delayed events, respectively. It can be seen from Fig.\ref{fig11} (a) for prompt events, the distribution has an abrupt jump in both data and MC events. From simulations we have seen that most of the times positron deposits its entire energy in a single PS bar~\cite{ISMRAN}. The annihilation $\gamma$-rays may span multiple bars. To reduce the singles $\gamma$-ray  reactor background contamination for the prompt events in the RON data, we use higher $\mathrm{N_{bars}}$ selection. One of the reasons for the jump, in the  $\mathrm{E_{max} / E_{sum}}$ for PS bars from prompt events, is due to the bias introduced in selection criteria applied to individual PS bars, $\mathrm{E^{ p}_{sum}}$ and $\mathrm{N_{bars}}$ while defining the prompt events in the data and MC. Also this jump is more pronounced in data as compared to the distribution from MC events. One of the reasons could be the reactor related background in data, which is not completely reproduced by the MC generated events in the prompt event candidates.
On contrary, for the delayed events, the cascade $\gamma$-ray distribution is rather isotropic and not concentrated in one single PS bar and hence reducing the bias on the event selection criteria. This can be seen from a reasonable comparison between data and MC events in Fig.~\ref{fig11} (b) for delayed events.
\begin{figure}[h]
\begin{center}
\includegraphics[scale=0.65]{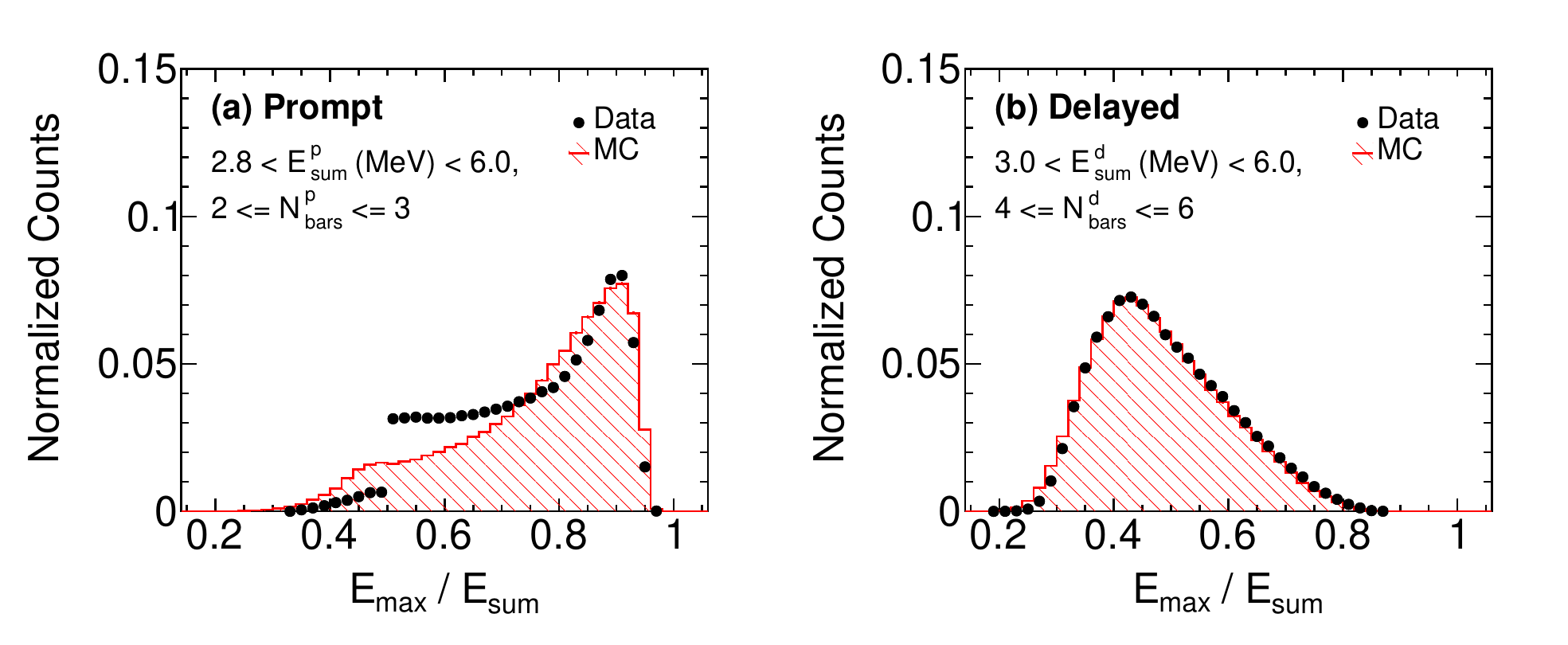}
\caption{Comparison between data (solid) and MC (shaded) for ratio of $\mathrm{E_{max} / E_{sum}}$ for (a) prompt and (b) delayed events.}
\label{fig11}
\end{center}
\end{figure}
It is also important to look at the energy sharing among the PS bars in prompt and delayed events. Figure~\ref{fig12} (a) and (b) show the $\mathrm{E_{1} / E_{max}}$ distribution for the prompt and delayed events, respectively. Again it can be seen, that for the prompt events, there is a disagreement between data and MC events as compared to that from the delayed events. Applying selection criteria on the $\mathrm{E_{max} / E_{sum}}$ and $\mathrm{E_{1} / E_{max}}$ ratios may improve the background rejection for the prompt and delayed events.
\begin{figure}[h]
\begin{center}
\includegraphics[scale=0.65]{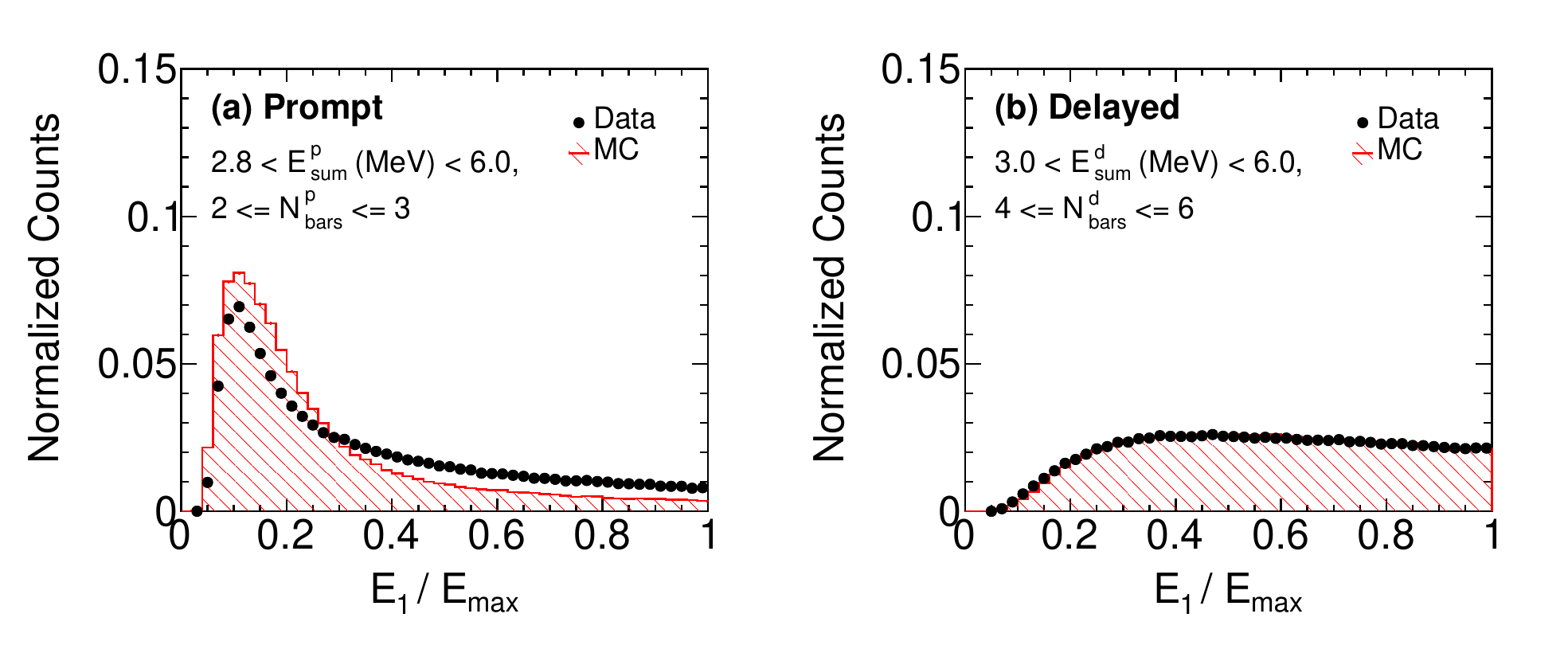}
\caption{Comparison between data (solid) and MC (shaded) for ratio of $\mathrm{E_{1} / E_{max}}$ in (a) prompt and (b) delayed events.}
\label{fig12}
\end{center}
\end{figure}

\begin{figure}[h]
\begin{center}
\includegraphics[scale=0.65]{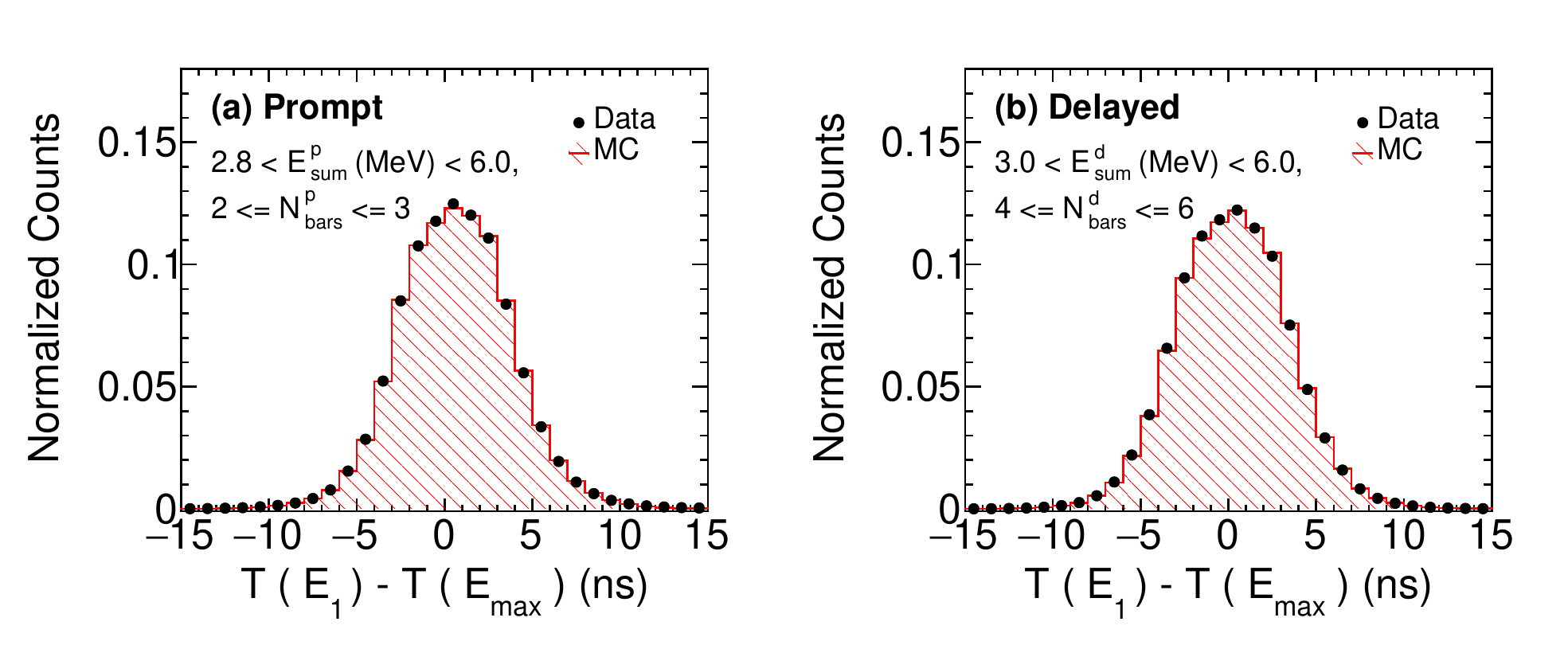}
\caption{Comparison between data (solid) and MC (shaded) for difference in timestamps for the PS bars with $\mathrm{E_{max}}$ and $\mathrm{E_{1}}$ in (a) prompt and (b) delayed events.}
\label{fig13}
\end{center}
\end{figure}
\begin{figure}[h]
\begin{center}
\includegraphics[scale=0.65]{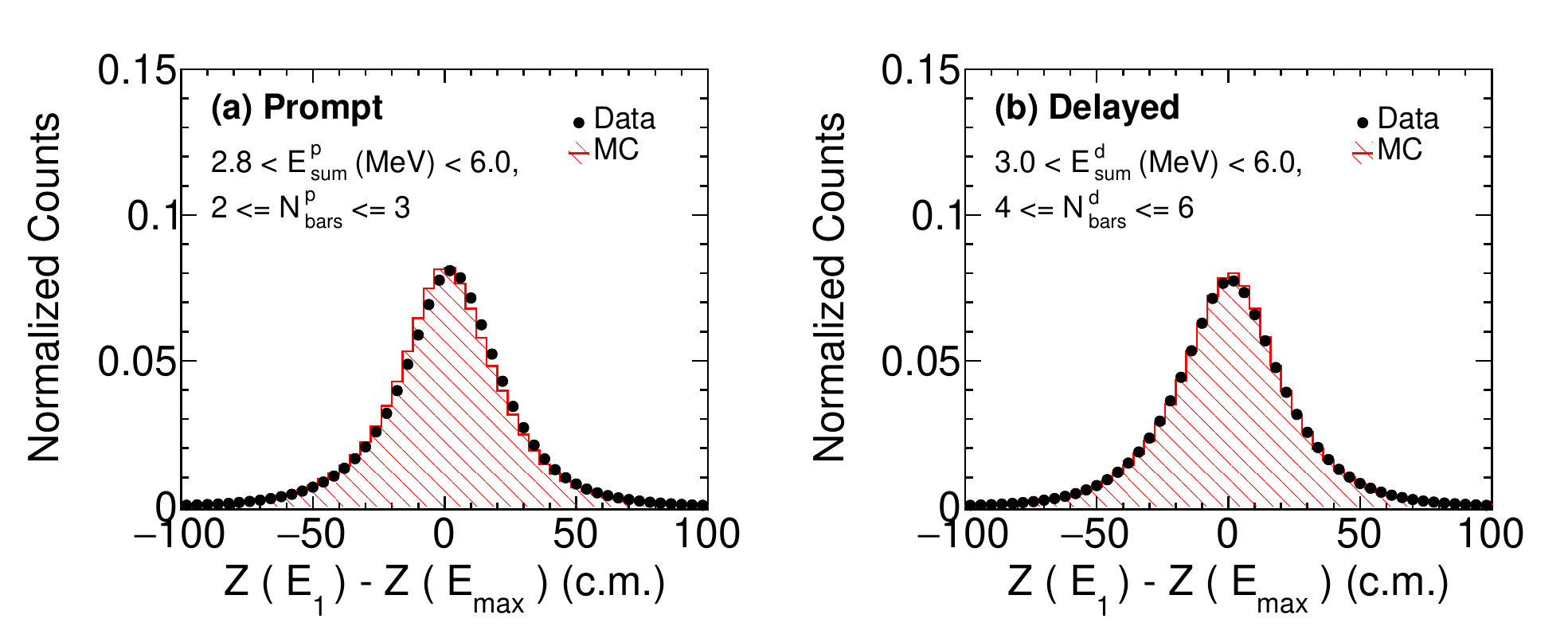}
\caption{Comparison between data (solid) and MC (shaded) for difference in $\mathrm{Zpos}$ for the PS bars with $\mathrm{E_{max}}$ and $\mathrm{E_{1}}$ in (a) prompt and (b) delayed events.}
\label{fig14}
\end{center}
\end{figure}
\begin{figure}[h]  
\begin{center}
\includegraphics[scale=0.35]{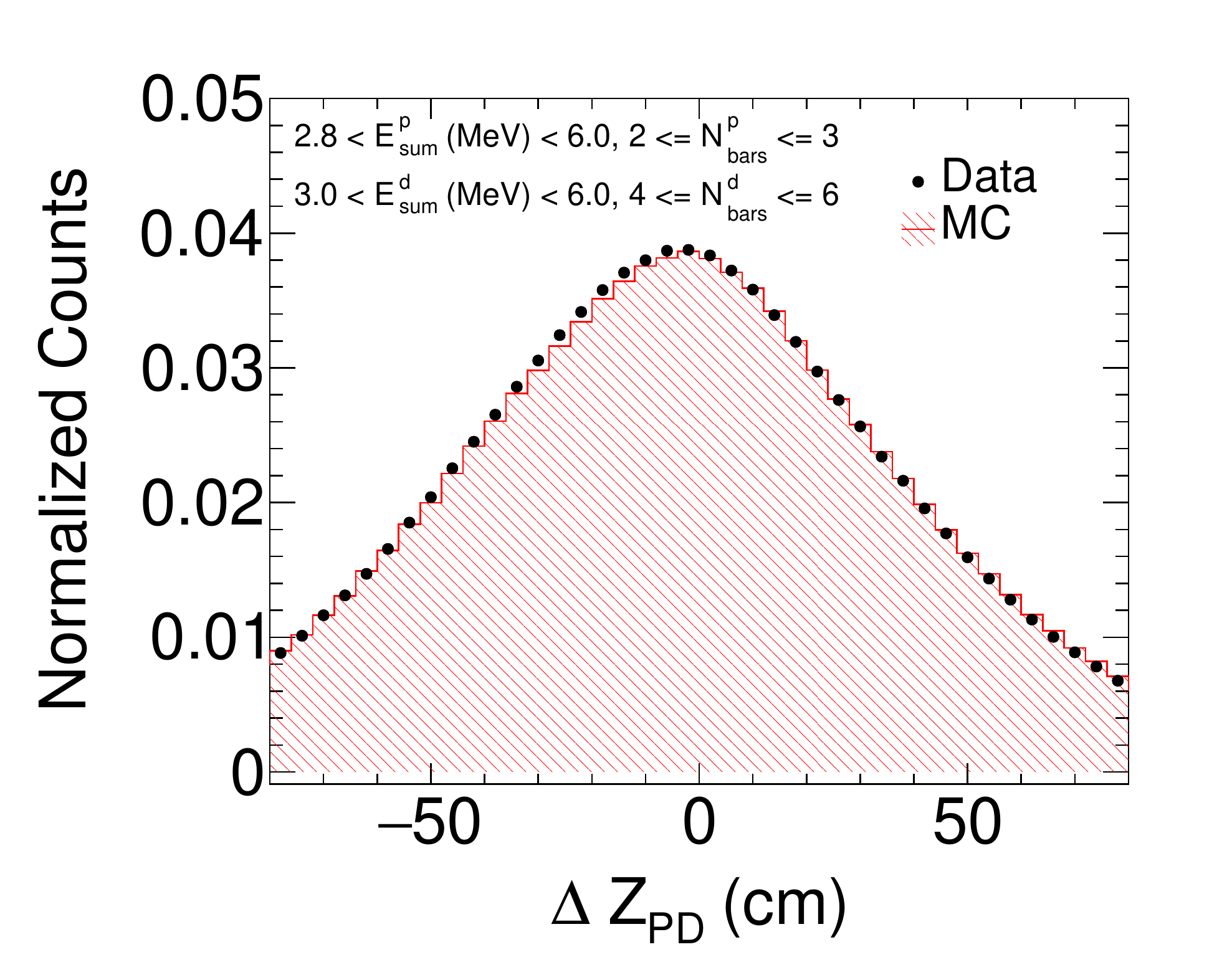}
\caption{Comparison between data (solid) and MC (shaded) for $\mathrm{\Delta Z_{PD}}$ between prompt and delayed events. The $\mathrm{Zpos}$ for prompt and delayed event are taken for the PS bar having $\mathrm{E_{max}}$.}
\label{fig15}
\end{center}
\end{figure}
\begin{figure}[h]
\begin{center}
\includegraphics[scale=0.65]{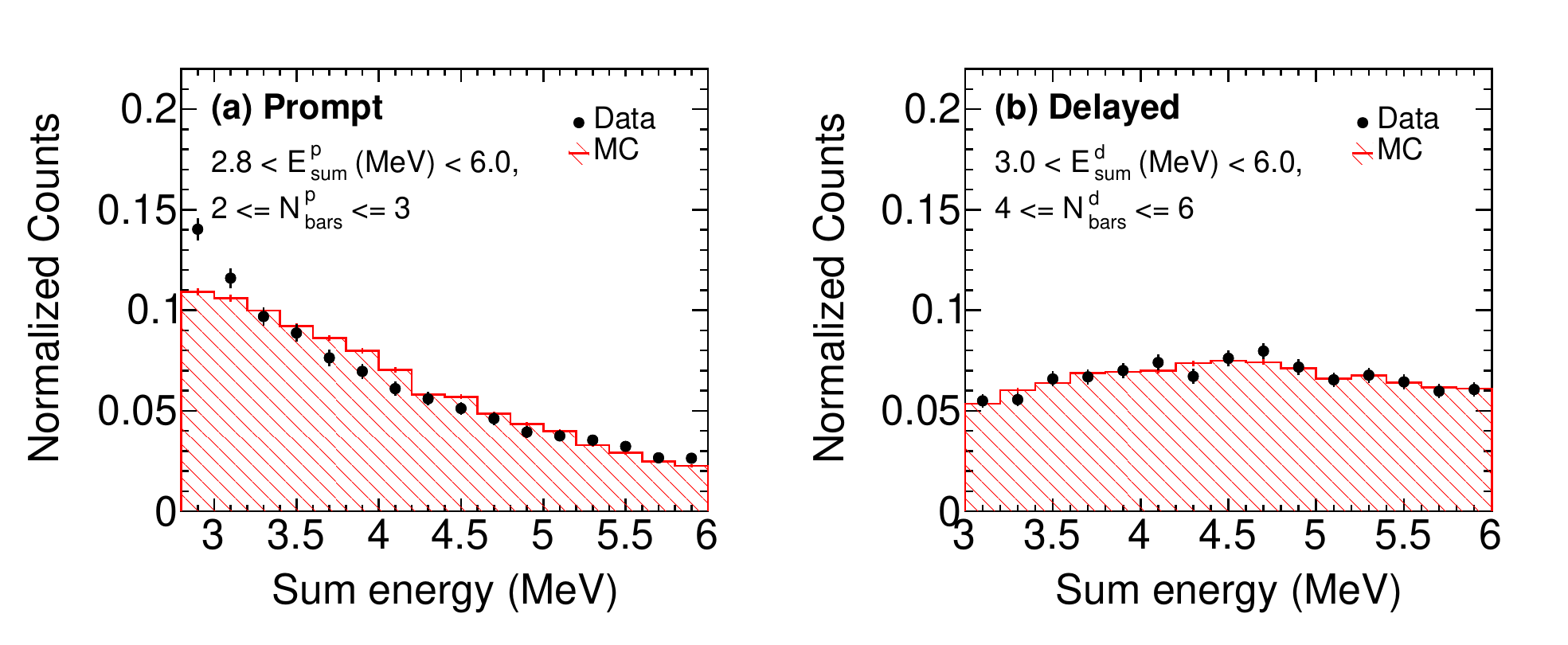}
\caption{Comparison between data (solid) and MC (shaded) for sum energy $\mathrm{E_{sum}}$ in (a) prompt and (b) delayed events.}
\label{fig16}
\end{center}
\end{figure}
\begin{figure}[h]
\begin{center}
\includegraphics[scale=0.65]{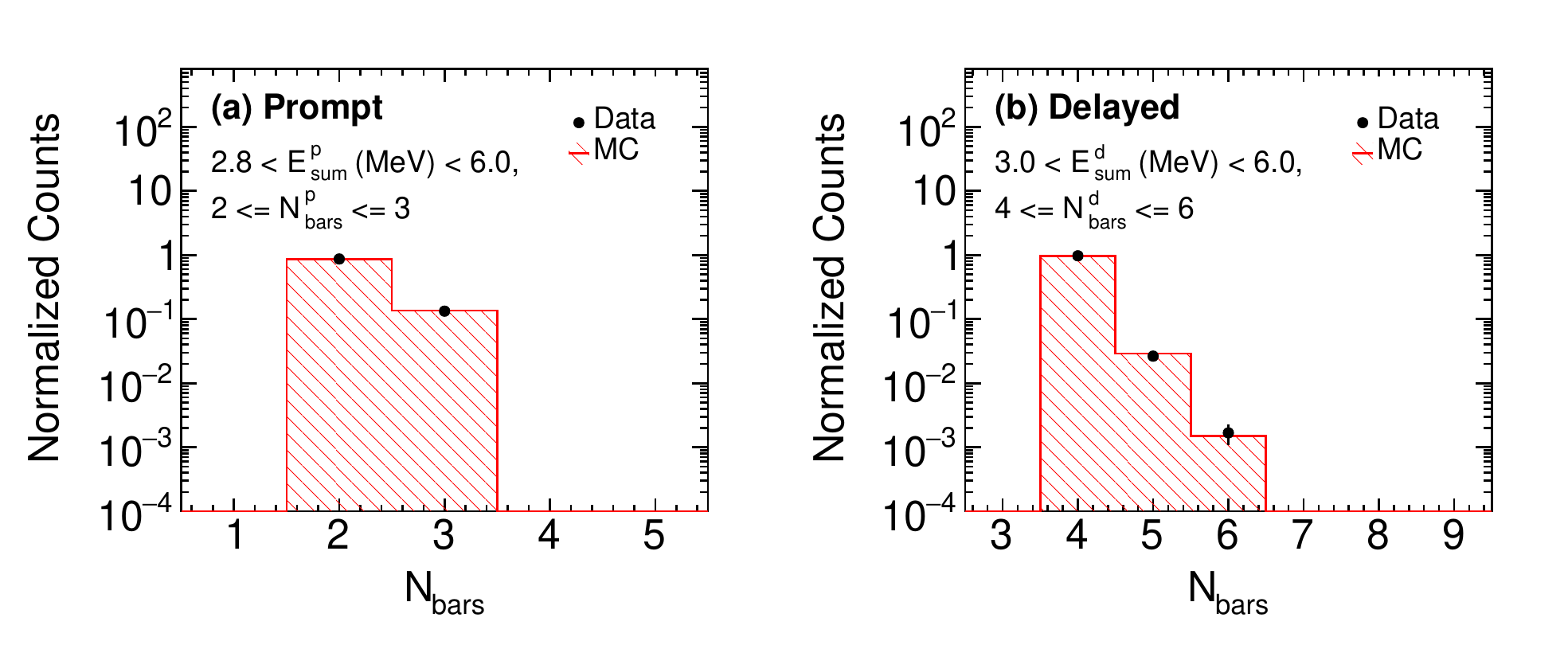}
\caption{Comparison between data (solid) and MC (shaded) for number of bars hit $\mathrm{N_{bars}}$ in (a) prompt and (b) delayed events.}
\label{fig17}
\end{center}
\end{figure}
Due to the finite segmentation of the PS bars in the mini-ISMRAN, we exploit the timestamp information recorded for the PS bars to study the spread of the events in time. The correlated events arising from prompt or delayed event should be close in time as opposed to the uncorrelated accidental background events. Figure~\ref{fig13} (a) and (b) show the difference in timestamp for prompt and delayed events in the PS bars having $\mathrm{E_{max}}$ and $\mathrm{E_{1}}$, respectively. To estimate the width of these distributions, we fit the prompt and delayed events time spread between $\mathrm{E_{max}}$ and $\mathrm{E_{1}}$ with a double Gaussian function. The time spread between the PS bars in prompt and delayed candidate events is found to be $\sim$3 ns. Similarly, the spread in the Zpos between the PS bars having $\mathrm{E_{max}}$ and $\mathrm{E_{1}}$ for prompt and delayed events is found to be $\sim$14 cm as shown in  Fig.~\ref{fig14} (a) and Fig.~\ref{fig14} (b), respectively.
%It can be seen from the distribution that most of the events lie within $\pm$ 5 ns. Also the time spread or width of the distribution for prompt bars is smaller as compared to that for the delayed bars. This can be understood in terms of the cascade $\gamma$-ray emission process in delayed event, which is mostly an isotropic distribution and can span larger number of PS bars as opposed to the prompt positron event, where most of the positron energy is deposited in a single PS bar, leaving the two annihilated $\gamma$-rays to span multiple bars.
%Similarly, the spread in the $\mathrm{Zpos}$ between the PS bars having $\mathrm{E_{max}}$ and $\mathrm{E_{1}}$ for prompt and delayed events can be seen from Fig.~\ref{fig14} (a) and Fig.~\ref{fig14} (b), respectively. 
\begin{figure}[h]
\begin{center}
\includegraphics[scale=0.70]{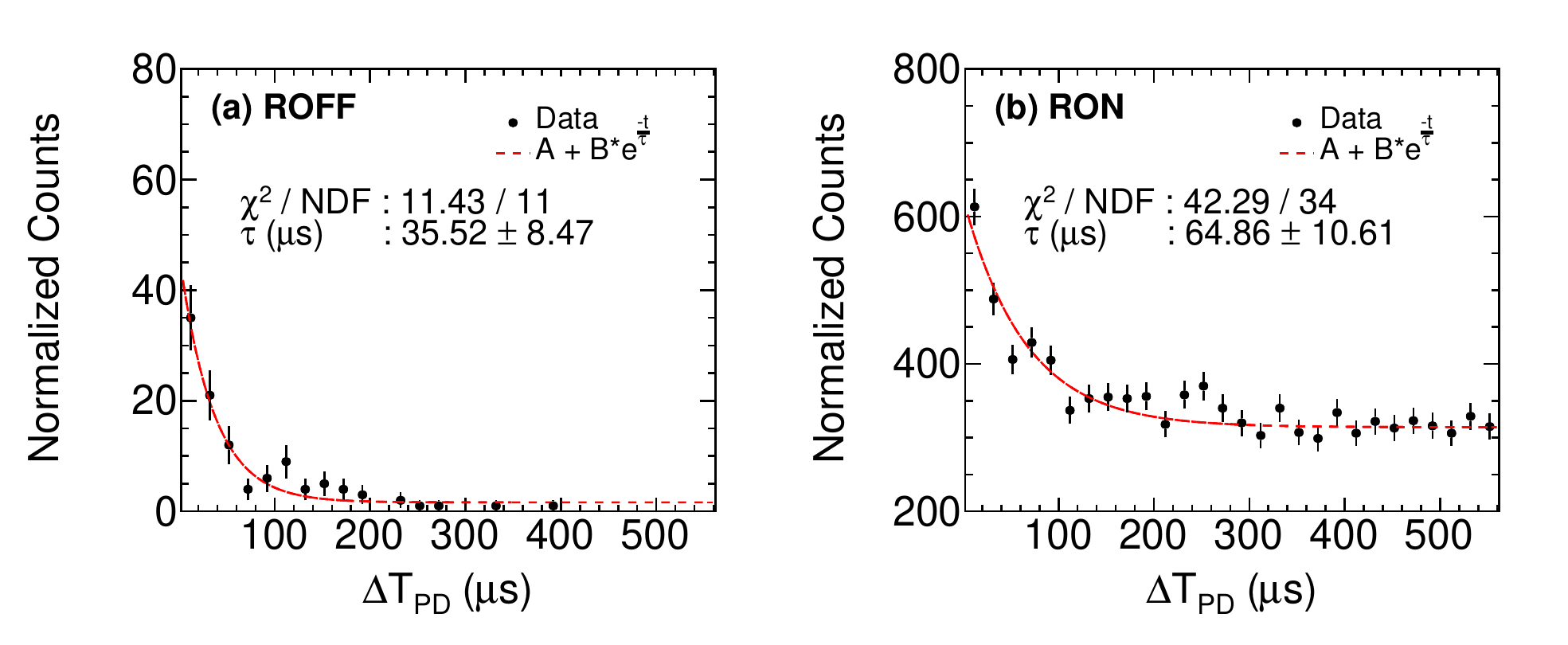}
\caption{Time difference ($\mathrm{\Delta T_{PD}}$) between prompt and delayed event pairs for (a) ROFF and (b) RON condition. Also shown are the fit functions with an exponential and a constant term. Errors shown on $\mathrm{\tau}$ are statistical and fit errors added in quadrature.}
\label{fig18}
\end{center}
\end{figure}
\begin{figure}[h]
\begin{center}
\includegraphics[scale=0.70]{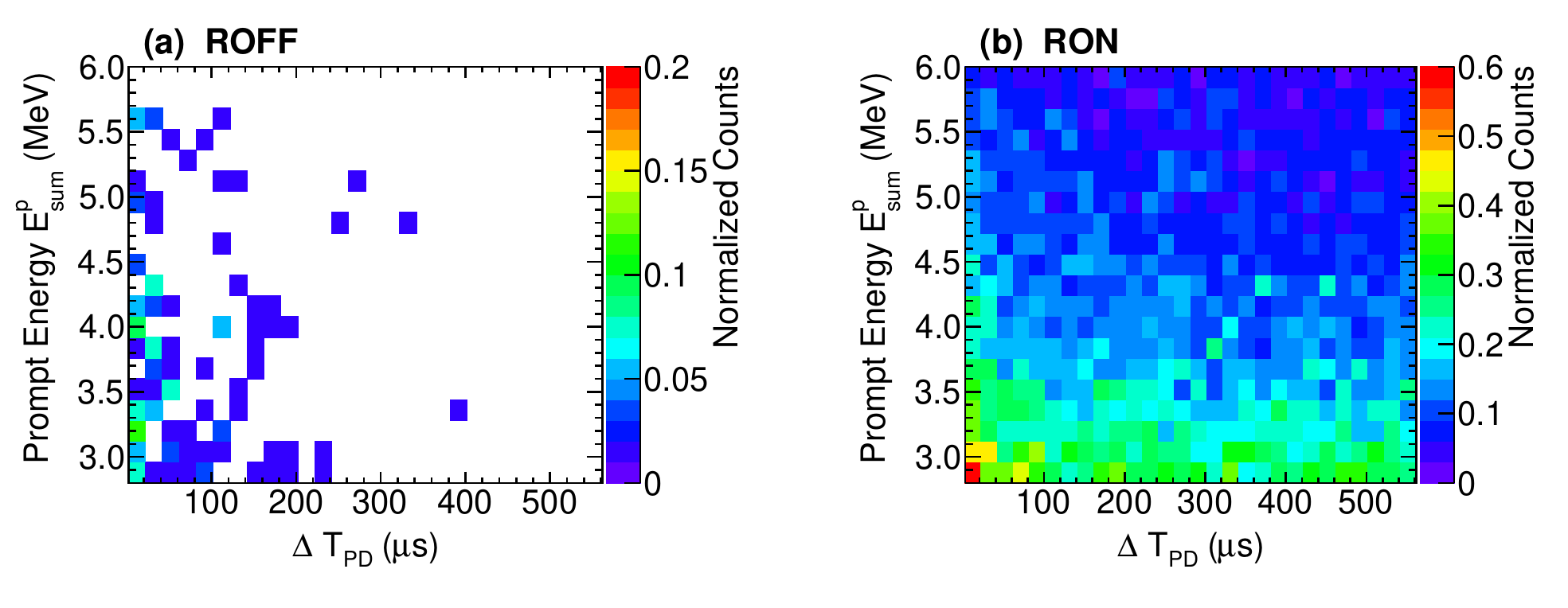}
\caption{Prompt sum energy ($\mathrm{E^{p}_{sum}}$) for candidate events as a function of time difference ($\mathrm{\Delta T_{PD}}$) for (a) ROFF and (b) RON condition.}
\label{fig18a}
\end{center}
\end{figure}

\begin{table}[h]
\begin{small}
  \begin{center}
  \caption{Detection efficiencies of $\overline\nuup_{e}$ events in mini-ISMRAN with different prompt and delayed event selections.}
  \label{table1}
\begin{tabular}{|c|c|}%|p{5.0cm}|p{1.25cm}|p{5.0cm}|p{1.25cm}|}
\hline
\makecell{Selection Criteria} & Efficiency ($\%$) \\
\hline
\makecell{$\mathrm{E_{sum}}$ and $\mathrm{N_{bars}}$} & 74 \\
\hline
\makecell{Muon rejection}  & 70  \\
\hline
\makecell{Fake prompts/delays}  & 61  \\
\hline
\makecell{$\mathrm{\Delta Z_{PD}}$ $<$ 1.50 $\sigma$}    & 52  \\
\hline
\makecell{$\mathrm{E^{p}_{max} / E^{p}_{sum}}$ $>$ 0.50}   & 44  \\
\hline
\makecell{$\mathrm{E^{p}_{1} / E^{p}_{max}}$ $<$ 0.40}     & 39 \\
\hline
\makecell{$\mathrm{E^{d}_{max} / E^{d}_{sum}}$ $>$ 0.25}   & 31  \\
\hline
\makecell{$\mathrm{E^{d}_{1} / E^{d}_{max}}$ $<$ 0.65}     & 23  \\
\hline
\makecell{Prompt/Delay time spread $<$ 1.5 $\sigma$}     & 12  \\
\hline
\makecell{Prompt/Delay position spread $<$ 1.5 $\sigma$} & 4  \\
\hline
\end{tabular}
\end{center}
\end{small}
\end{table}
In both cases, timing and position, there is a reasonable agreement between the data and MC events. We fit these distributions with a double Gaussian function and apply a selection criteria on the timing and Z position in the prompt and delayed events, which are within 1.5$\sigma$ of the distributions.

Since in mini-ISMRAN, we can reconstruct the position of the event along the length axis of the PS bar, we can use this position information to look for the position spread between the prompt and the delayed events. Figure~\ref{fig15} shows the difference of the parameterized Z position obtained from timing ($\mathrm{\Delta T_{LR}}$) for the PS bar with $\mathrm{E_{max}}$ from the prompt and delayed event. We use this information to further restrict the selection of prompt and delayed events which fall within 1.5 $\sigma$ of this distribution.
After applying all these selection criteria on prompt and delayed events, we look at the $\mathrm{E_{sum}}$ and $\mathrm{N_{bars}}$ distribution for prompt and delayed events. 
Figure~\ref{fig16} (a) and (b) show the $\mathrm{E_{sum}}$ energy distribution for the prompt and delayed events, respectively. The results are compared with the MC simulations. For the prompt events, below $\mathrm{E^{p}_{sum}}$ $<$ 3.2 MeV, there is a disagreement between data and MC events. This disagreement may be due to differences in the data and MC events for individual PS bar energy spectra for the prompt events. Other reasons may be due to presence of residual background events on which the above mentioned selection criteria are less effective. For delayed events, as shown in Fig.~\ref{fig16} (b),  the agreement between data and MC events is good. Due to limited acceptance of the mini-ISMRAN detector, the complete reconstruction of cascade $\gamma$-rays is not possible and does not allow us to see a peak structure at $\sim$8MeV.
Figure~\ref{fig17} (a) and (b) show the $\mathrm{N_{bar}}$ distribution for prompt and delayed events, respectively. There is a very good agreement between data and MC events for both prompt and delayed events.
The time difference ($\mathrm{\Delta T_{PD}}$) between prompt and delayed event pairs is an important variable to estimate the $\overline\nuup_{e}$ events. Figure~\ref{fig18} (a) and (b) show the $\mathrm{\Delta T_{PD}}$ distribution for ROFF and RON conditions, respectively. The $\mathrm{\Delta T_{PD}}$ distributions are time normalized and are fitted to a combined function consisting of an exponential term for the neutron thermalization and capture time in PS bars and a constant term representing the accidental residual background. For RON, the fit results in a characteristic time ( $\mathrm{\tau}$ ) of 64.86 $\mathrm{\mu}$s $\pm$ 10.16 $\mathrm{\mu}$s and is in good agreement with $\sim$68 $\mathrm{\mu}$s obtained from MC events~\cite{ISMRAN}. This again ensures that the events which are filtered after the selection criteria are mainly from the neutron capture on Gd and most probably consists of ${\overline{\ensuremath{\nu}}}_{e}$ candidate events. For ROFF, the $\mathrm{\tau}$ between prompt and delayed event pairs is 35.52 $\mathrm{\mu}$s $\pm$ 8.47 $\mathrm{\mu}$s which is mainly due to correlated reactor or cosmogenic background events. This can be attributed to the decay of long lived isotopes or from the residual activities from the refueling of the reactor. Figure~\ref{fig18a} (a) and (b) show the prompt energy variation as a function of $\mathrm{\Delta T_{PD}}$ for ROFF and RON conditions, respectively. It can be seen for RON, from Fig.~\ref{fig18a} (b), the events corresponding to $\mathrm{\Delta T_{PD}}$ $>$ 250 $\mu$s are almost unifromly distributed along the $\mathrm{\Delta T_{PD}}$ and are with $\mathrm{E^{p}_{sum}}$ $\sim$ 3.0 MeV. This feature indicates the dominant contribution to the $\mathrm{E^{p}_{sum}}$ at larger $\mathrm{\Delta T_{PD}}$ originates from the reactor related background which can be also be seen in Fig.~\ref{fig2} from the singles $\gamma$-ray spectra measured using $\mathrm{CeBr_{3}}$ detector. In ROFF, shown in  Fig.~\ref{fig18a} (a), no such feature is observed within the measured statistics.

The comparative results between the real data and MC events gives us the confidence to apply the selection criteria on the prompt and delayed events in real data and to calculate the ${\overline{\ensuremath{\nu}}}_{e}$ reconstruction efficiency from mini-ISMRAN using embedded MC events.
For the reconstruction efficiency calculations, only those events are considered where the neutron is captured on Gd foils. Since with neutron capture on H leaves with 2.2 MeV mono-energetic $\gamma$-ray and most of the times would not fit in our selection criteria of the delayed event. 
A detailed Table~\ref{table1} showing the various selection criteria applied on the prompt and delayed events and their effect on the ${\overline{\ensuremath{\nu}}}_{e}$ reconstruction efficiency in mini-ISMRAN detector. The successive reduction in efficiency is shown for the various selection criteria applied on the prompt and delayed events. An efficiency of $\sim$4 $\%$ is achieved for the mini-ISMRAN from the embedding technique using the MC events. A systematic uncertainty of $\sim$6 $\%$ is estimated on the efficiency by varying different event selection criteria. Though the uncertainties on the efficiency are large at the moment and mainly because of the smaller acceptance of the mini-ISMRAN. The final errors for the reconstruction efficiencies in the full scale ISMRAN will be much smaller.
There can be other sources of background events like fast neutron background event, which is an irreducible background for  ${\overline{\ensuremath{\nu}}}_{e}$ events. The proton recoil from the fast neutron mimics the prompt event signature and the capture of the same neutron after thermalization in one of the PS bars provides the delayed event signature. To estimate the efficiency of this particular background, we have generated an uniform distribution of the fast neutron in energy range from 2 to 12 MeV and passed them through GEANT4 for the response on the PS bars. We again use the same embedding technique to estimate their detection efficiency in the mini-ISMRAN. We have applied same selection criteria which are applied on filtering of the prompt and delayed events from ${\overline{\ensuremath{\nu}}}_{e}$ spectra, on the fast neutron events. We obtained an efficiency of $\sim$0.3$\%$ with our event selection criteria for the fast neutrons which can mimic ${\overline{\ensuremath{\nu}}}_{e}$ event signatures in mini-ISMRAN. This is added as an additional systematic uncertainty in background subtraction in our final results. Similar study is done with the cosmogenic muon induced $\mathrm{{}^{12}}$B events~\cite{B12} and it is found that their efficiencies in mini-ISMRAN is even less than $\sim$0.01$\%$ with our event selection criteria.
The systematic uncertainties arising from different selection criteria based on energy, time and position are conservatively estimated for the mini-ISMRAN. The time and position uncertainties are correlated, similarly the uncertainties from energy selection are also reflected in the energy ratios. The coincidence timing uncertainity $\mathrm{\Delta T_{LR}}$ is estimated by varying the coincidence window for individual PS bar by 5 ns. For the estimation of the uncertainties in the Zpos of the PS bars in selecting the prompt and delayed candidates is done by selecting the PS bars with $\pm$ 30 cm and $\pm$ 40 cm. The uncertainty in the Zpos difference between prompt and delayed $\mathrm{\Delta Z_{PD}}$ candidate events are estimated by varying the selection criteria by 1.5 sigma to 2.0 sigma. The systematic uncertainty on the energy ratio cuts is estimated by varying the selection criteria by 5$\%$ on the individual ratios, $\mathrm{E_{max}/E_{sum}}$ and $\mathrm{E_{1}/E_{max}}$, separately.
The uncertainty in the efficiency estimation using embedding technique is estimated by varying the timing, position and energy resolution by 5$\%$. The estimation of background subtraction in different regions of $\Delta T_{PD}$ is estimated by varying the fit range from 350 to 500 $\mu$s.

Table~\ref{table2} shows the systematic uncertainties from different variables on the final $\overline\nuup_{e}$ rates. The uncertainty in energy measurement in each PS bar is treated separately and not added in the $\mathrm{E_{sum}}$ for the prompt and delayed events.
\begin{table}[h]
\begin{small}
  \begin{center}
  \caption{Systematic uncertainties for mini-ISMRAN for the estimation of $\overline\nuup_{e}$ events.}
  \label{table2}
\begin{tabular}{|c|c|}%|p{5.0cm}|p{1.25cm}|p{5.0cm}|p{1.25cm}|}
\hline
\makecell{Selection Criteria} & Systematic uncertainty ($\%$) \\
\hline
\makecell{$\mathrm{\Delta T_{LR}}$}   & 7.3 \\
\hline
\makecell{$\mathrm{\Delta Z_{PD}}$}   & 3.9 \\
\hline
\makecell{Energy ratio cuts}         & 7.1  \\
\hline
\makecell{Timing cuts}               & 1.8  \\
\hline
\makecell{Z position cuts}           & 5.3  \\
\hline
\makecell{efficiency estimation}     & 4.0  \\
\hline
\makecell{Background subtraction}    & 12.0 \\
\hline
\end{tabular}

\end{center}
\end{small}
\end{table}
A total of $\sim$ 17 $\%$ of systematic uncertainty is estimated for the mini-ISMRAN in RON condition in Dhruva reactor hall. Some of these errors will improve in the final ISMRAN setup, due to the larger acceptance of the events and better fiducialization of the detector volume.
\section{Results and discussion}
With the selection of the prompt and delayed candidate events, we now discuss the methods for obtaining $\overline\nuup_{e}$ candidate events. A total of 128 days of RON and 51 days of ROFF data is analyzed. We follow two fold approach to cross check the stability of our results from the mini-ISMRAN. 
\begin{figure}[h]
\begin{center}
\includegraphics[scale=0.70]{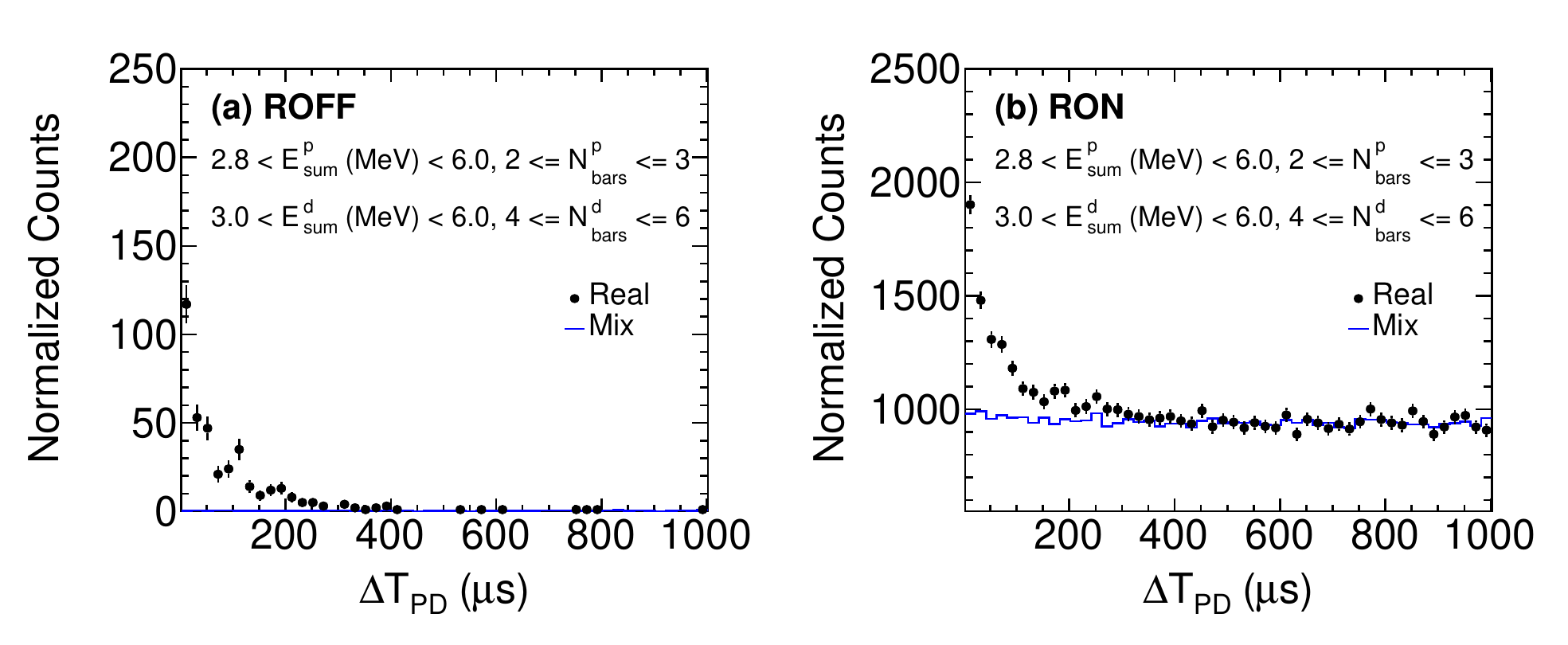}
\caption{Comparison of $\mathrm{\Delta T_{PD}}$ for real (solid) and mixed data (line) for ROFF and RON condition.}
\label{fig19}
\end{center}
\end{figure}
Figure~\ref{fig19} (a) and (b) show the time difference ($\mathrm{\Delta T_{PD}}$) distribution between prompt and delayed events for ROFF and RON conditions, respectively. The distribution shown in the Fig.~\ref{fig19} is obtained with prompt and delayed event pairs with $\mathrm{E_{sum}}$, $\mathrm{N_{bars}}$, $\mathrm{Zpos} $and time difference selection criteria for the demonstration purpose. For the final estimation of $\overline\nuup_{e}$ candidate events all the selection criteria discussed in the above section are applied to the reconstructed prompt and delayed events.
The black solid points represents the $\mathrm{\Delta T_{PD}}$ for all the prompt-delayed pairs reconstructed from real data events within 1000 $\mathrm{\mu}$s. Also shown, in continuous blue line, $\mathrm{\Delta T_{PD}}$ obtained from a data driven ``mixed event'' technique which is used to estimate the accidental background contribution~\cite{MixedEvent}. In this technique we use all reconstructed prompt events from real data and shift their respective timestamps by 1 ms. This is done to ensure that any correlation of signal or background event are explicitly broken between the prompt-delayed pairs. The shifting procedure is done consecutively 5 times for each prompt candidate event to get a reasonable statistics for the mixed data events. Once we have the shifted the prompt event in time, we follow the whole analysis procedure, including the event selection criteria, to again reconstruct the prompt-delayed pairs to obtain the mixed data events. Figure~\ref{fig19} (a) for ROFF, the prompt-delayed pairs constructed from real data events are either from the accidental background or from the cosmogenic muon induced events. This correlation is explicitly broken and can be seen in the prompt-delayed pairs from mixed data events. Figure~\ref{fig19} (b) shows the correlation between prompt-delayed pairs in time for the real and mixed data events. Above $\mathrm{\Delta T_{PD}}$ $>$ 350 $\mathrm{\mu}$s there is scaling in real and mixed data events, indicating the purely accidental component arising from background events in mini-ISMRAN. The mixed event technique not only defines the scale of the pure accidental background in the real data, but also helps in circumventing the uncertainties in the background subtraction in real data due to the large statistical uncertainties in the background region in the current measurements.
To obtain the $\overline\nuup_{e}$ candidate events, we divide $\mathrm{\Delta T_{PD}}$ distribution in two regions. One region consisting of prompt-delayed pairs within 8 $\mathrm{\mu}$s $<$ $\mathrm{\Delta T_{PD}}$ $<$ 250 $\mathrm{\mu}$s denoted as the signal with background (S+B) events and secondly those prompt-delayed pairs within 350 $\mathrm{\mu}$s $<$ $\mathrm{\Delta T_{PD}}$ $<$ 550 $\mathrm{\mu}$s which are purely dominated by accidental background (B) events in mini-ISMRAN. The $\overline\nuup_{e}$ candidate events in real data are then obtained by subtracting the integrated yield in both the regions by applying the appropriate normalizing factor to the pure accidental background (B) events. This we denote as method I. In this method, it is assumed that the background level is constant and uniform over the entire $\mathrm{\Delta T_{PD}}$ region. This may not be true as the accidental background is correlated with reactor power.

To cross-check our results obtained in the above manner, we also use the mixed data events (from blue continuous line) to perform the background subtraction of the prompt-delayed pairs from the real data events. This we denote as method II. As shown from Fig.~\ref{fig19} (b) blue continuous line, we can construct the pure accidental prompt-delayed background pairs in the S+B region using the mixed data. The $\mathrm{\Delta T_{PD}}$ distribution obtained from mixed data are normalized such that the subtraction of the integrated yield of prompt-delayed pairs in the B region for real and mixed data is zero. Then the integrated yield of the prompt-delayed pairs in S+B region from the mixed data distribution is taken as the accidental background.
From real data the prompt-delayed pairs obtained in the S+B region is 1361 $\pm$ 37. The prompt-delayed background pairs for RON condition from method I (i.e. integrated yield from B region in real data) and method II (i.e. integrated yield from S+B region in mixed data) are 1093 $\pm$ 34 and 1118 $\pm$ 17, respectively. Similarly for the ROFF condition, from the real data we obtain 21 $\pm$ 5 prompt-delayed pairs in S+B region and 1 $\pm$ 1 prompt-delayed pairs from both the background methods. The errors quoted here are only statistical in nature. 
The $\overline\nuup_{e}$ candidate events obtained in RON from the subtraction method I are 268 $\pm$ 50 and from method II 243 $\pm$ 41. The results are consistent within the statistical errors and confirms the selection of signal and background regions are reasonable. For ROFF condition, the accidental background obtained from the above two methods yields 20 $\pm$ 5. Again, all errors quoted are only statistical.

To obtain final $\overline\nuup_{e}$ yield, using method I and II, we use the prescription defined in the Ref.~\cite{ANPROSPECT}.
\begin{equation}\label{eq:an1}
  \hspace{-0.2in}
   \mathrm{N^{I}_{\overline\nuup_{e}}~(IBD) = [~(~N^{R}_{RON}~(S+B) - N^{R}_{RON}~(B)~) - K~(~N^{ R}_{ROFF}~(S+B) - N^{R}_{ROFF}~(B)~)}~],
\end{equation}
\begin{equation}\label{eq:an2}
  \hspace{-0.2in}
   \mathrm{N^{II}_{\overline\nuup_{e}}~(IBD) = [~(~N^{R}_{RON}~(S+B) - N^{M}_{RON}~(S+B)~) - K~(~N^{R}_{ROFF}~(S+B) - N^{M}_{ROFF}~(S+B)~)}~],
\end{equation}
where,  $\mathrm{N^{ R}_{RON(ROFF)}}$ and $\mathrm{N^{ M}_{RON(ROFF)}}$ are the integrated yields from the real data and mixed data for S+B and B regions for either RON or ROFF condition, respectively. $\mathrm{N^{I}_{\overline\nuup_{e}}}$ and $\mathrm{N^{II}_{\overline\nuup_{e}}}$ from Eqs.~\ref{eq:an1} and ~\ref{eq:an2} define the $\overline\nuup_{e}$ yield obtained from method I and II, respectively. The scaling factor K represents the ratio of number of RON to ROFF days. A $\sim$12 $\%$ systematic uncertainty, as shown in Table~\ref{table2}, on the background subtraction methods is estimated by varying the background regions in $\mathrm{\Delta T_{PD}}$ distribution. The total systematic uncertainties obtained for the final $\overline\nuup_{e}$ events is 17$\%$. 

Using the above mentioned prescription a total of 218 $\pm$ 50 (stat) $\pm$ 37 (sys) from method I and 192 $\pm$ 41 (stat) $\pm$ 33 (sys) from method II~$\overline\nuup_{e}$ events in mini-ISMRAN are obtained for the period of 128 days of RON condition. 
We compare our results with the theoretical estimation which is obtained using the equation from Ref.~\cite{VVER}. The uncertainties in estimation of the predicted rate mainly comes from the reactor power, fiducial volume, reconstruction efficiency, flux uncertainties and distance from the reactor core. Moreover, the effect of fission yields on $\overline\nuup_{e}$ spectra and the contributions from the top three isotopes to the spectra may arise uncertainties in the flux calculations~\cite{Sonzogni,Rasco}. A conservative estimate of 10$\%$ of uncertainties are quoted on the expected rate for $\overline\nuup_{e}$ for mini-ISMRAN. We have taken the efficiency for the $\overline\nuup_{e}$ events from embedding as 4$\%$ and the estimation is done by taking the appropriate reactor power for the running time period. The expected $\overline\nuup_{e}$ event rate in mini-ISMRAN for RON condition for 128 days is 214 $\pm$ 32 which is in reasonable agreement with the measured data.

\section{Conclusions and Outlook}
We present measurements of a prototype array consisting of 16 plastic scintillator bars, mini-ISMRAN inside Dhruva reactor hall. The detector has been placed at $\sim$13 m from the reactor core enclosed in a passive shielding of 10 cm of lead and 10 cm of borated polyethylene. A total of 128 days and 51 days of reactor on and reactor off data is analyzed, respectively. A data driven technique, which consists of embedding of Monte Carlo generated events in real background from data is used to estimate the reconstruction efficiencies, rejection of muon induced events and fake rate for prompt and delayed events. A detailed comparison of data and MC events is presented for different selection variables, consisting of energy, time and position in PS bars from mini-ISMRAN, for both prompt and delayed events. Depending on these selection criteria, prompt and delayed candidate events are selected to estimate the $\overline\nuup_{e}$ candidate events. Two different approaches are compared for the background subtraction, with real and mixed prompt-delayed pairs, to obtain $\overline\nuup_{e}$ candidate events. For reactor on condition 268 $\pm$ 50 and 243 $\pm$ 41 $\overline\nuup_{e}$ candidate events are obtained for two background subtraction methods, respectively.
For reactor off condition a total of 20 $\pm$ 5 (stat) background events are reconstructed which pass all the selection criteria for the $\overline\nuup_{e}$ candidate events. By taking into account the background events from ROFF condition and performing a scaled subtraction from the $\overline\nuup_{e}$ candidate events from reactor on condition, a total of 218 $\pm$ 50 (stat) $\pm$ 37 (sys) and 192 $\pm$ 41 (stat) $\pm$ 33 (sys) $\overline\nuup_{e}$ events from real and mixed data background subtraction are obtained, respectively. The full scale ISMRAN experiment is currently been installed in the Dhruva reactor hall on a movable base structure. Arrangement of the interlocked passive shielding of 10 cm of lead and 10 cm of borated polyethylene are been mounted on the base structure. The final detector assembly will be done in September 2021 and the physics data operations will begin from mid October 2021.
\section{Acknowledgments}
We are thankful to Reactor Operations Division (ROD) and Research Reactor Services Division (RSSD) and EmA$\&$ID workshop for logistic support and co-ordination, Centre for Design and Manufacture (CDM), BARC for taking up design and fabrication of the final ISMRAN support structure. We would like to thank S.~Bhatt, V.~R.~Gumma from CDM, BARC for the designing base and shield structures for ISMRAN. We thank Dr. U.~K.~Pal, NPD BARC for providing us the Cerium Bromide detector for the background measurements. We thank Dr. A.~Mitra, NPD BARC and C.~Tintori, CAEN for the helping in setting up of the data acquisition system for the mini-ISMRAN. We thank Dr. V.~K.~S.~Kashyap, NISER, Khurda Road for his efforts and discussions during his Ph.D period in the early conceiving stages of ISMRAN experiment. We would also like to thank Prof. V.~M.~Datar for the discussions during the intial stages of ISMRAN. A special thanks to Prof. A.~K.~Mohanty for very helpful discussions and giving all the support to ISMRAN experimental program.

\bibliographystyle{unsrt}  

%\bibliography{references}
\end{document}